\documentclass[%
 reprint,
superscriptaddress,
 amsmath,amssymb,
 aps,
]{revtex4-2}

 \usepackage[T1]{fontenc} 
 
\usepackage{graphicx}
\usepackage{dcolumn}
\usepackage{placeins}
\usepackage{bm}
\usepackage{lipsum}
\usepackage{braket}
\usepackage{siunitx}
\usepackage{epigraph} 
\usepackage{todonotes}
\usepackage{amsmath}
\usepackage{subfigure}
\usepackage{subfiles}

\usepackage{algorithm}
\usepackage{algpseudocode}

\DeclareMathOperator{\Tr}{Tr}

\begin{document}

\preprint{APS/123-QED}

\title{Learning Lindblad Dynamics of a Superconducting Quantum Processor}

\author{Johann Bock Severin}
\email{johann.severin@nbi.ku.dk}
\affiliation{NNF Quantum Computing Programme, Niels Bohr Institute, University of Copenhagen, Denmark}

\author{Malthe A. Marciniak}
\affiliation{NNF Quantum Computing Programme, Niels Bohr Institute, University of Copenhagen, Denmark}

\author{Rune Thinggaard Birke}
\affiliation{NNF Quantum Computing Programme, Niels Bohr Institute, University of Copenhagen, Denmark}
\affiliation{Center for the Mathematics of Quantum Theory, MATH Department, University of Copenhagen}

\author{Emil Hogedal}
\author{Andreas Nylander}
\author{Irshad Ahmad}
\author{Amr Osman}
\author{Janka Biznárová}
\author{Marcus Rommel}
\author{Anita Fadavi Roudsari}
\author{Jonas Bylander}
\author{Giovanna Tancredi}
\affiliation{Department of Microtechnology and Nanoscience, Chalmers University of Technology, SE-412 96 Gothenburg, Sweden}

\author{Christopher W. Warren}
\affiliation{NNF Quantum Computing Programme, Niels Bohr Institute, University of Copenhagen, Denmark}

\author{Svend Krøjer}
\affiliation{NNF Quantum Computing Programme, Niels Bohr Institute, University of Copenhagen, Denmark}

\author{Jacob Hastrup}
\affiliation{NNF Quantum Computing Programme, Niels Bohr Institute, University of Copenhagen, Denmark}

\author{Morten Kjaergaard}
\email{mkjaergaard@nbi.ku.dk}
\affiliation{NNF Quantum Computing Programme, Niels Bohr Institute, University of Copenhagen, Denmark}

\date{\today}

\begin{abstract}
Accurate models of quantum processors are essential for understanding, calibrating, and improving their performance. 
In practice, model construction must balance physical detail against the experimental and computational effort required to reliably learn parameters. 
Compact descriptions therefore often rely on assumptions about which interactions, noise processes, or hidden degrees of freedom are relevant. 
Here we introduce LIMINAL, a data-driven framework for testing such assumptions and selecting minimal adequate Lindblad models. 
LIMINAL fits nested candidate models to time-resolved tomographic data and uses likelihood-ratio tests to decide when added physical mechanisms are warranted. 
We apply LIMINAL to a five-qubit superconducting processor, identifying an idling model with three-local Hamiltonian terms and two-local dissipation, while finding no support for three-local dissipation. 
We further apply it to recover driven single-qubit Hamiltonians, reconstruct a shaped-pulse Hamiltonian without assuming an analytic pulse model, and test hidden-qubit extensions in coupler-mediated dynamics, demonstrating the applicability of the framework for a wide range of tasks.

\end{abstract}

\maketitle

\section{Introduction}
Accurate modeling of the dynamics and noise channels of a quantum processor is essential for calibration, error mitigation, and hardware development \cite{siddiqiEngineeringHighcoherenceSuperconducting2021, wittlerIntegratedToolSet2021, naeijOpenQuantumSystem2025}. In practice, however, developing suitable models faces a recurring trade-off: Models must be expressive enough to capture relevant physical effects, yet sufficiently constrained that their parameters can be estimated reliably without excessive experimental overhead \cite{hashimPracticalIntroductionBenchmarking2025a, briegerCompressiveGateSet2023a, vinasMicroscopicParametrizationsGate2025}. This, in turn, suggests the need for methods that identify the minimally adequate description of observable processor dynamics.

Quantum processor dynamics are typically described by a Lindblad master equation. Over a fixed time interval, the dynamics define a completely positive trace-preserving (CPTP) map that can, in principle, be reconstructed via process tomography \cite{poyatosCompleteCharacterizationQuantum1997, kneeQuantumProcessTomography2018}. Time-resolved variants enable estimation of a Lindblad generator under Markovian assumptions \cite{samachLindbladTomographySuperconducting2022a, varonaLindbladlikeQuantumTomography2025b, wallaceLearningDynamicsMarkovian2025}. In practice, however, both the number of Lindblad parameters and the number of tomographic configurations scale exponentially with qubit count, rendering unconstrained tomography infeasible \cite{cramerEfficientQuantumState2010, nielsenGateSetTomography2021a}.

To address this bottleneck, a wide range of scalable learning protocols impose physics-inspired structure, such as locality, sparsity, or restricted interaction graphs, so that estimation remains tractable at larger scales \cite{stilckfrancaEfficientRobustEstimation2024, hangleiterRobustlyLearningHamiltonian2024b, liuOptimalRobustInsitu2025a, birkeDemonstratingBenchmarkingClassical2026}. However, these structural assumptions are often adopted \emph{a priori}. Since device-specific couplings and hidden degrees of freedom can vary across devices and operating conditions, it is important to test whether a proposed reduced model is supported by experimental data, to determine when increasing model complexity ceases to provide explanatory value, and thereby to establish which physical assumptions are warranted by the data.

In this work, we address this issue by introducing a data-driven framework for model selection of quantum processor dynamics, which we term \emph{LIMINAL} (Likelihood-based Inference of Minimal Adequate Lindblad models). LIMINAL works by evaluating nested candidate Lindblad models using maximum-likelihood estimation (MLE) of time-resolved measurement data. The models are then compared using likelihood-ratio (LR) tests grounded in Wilks’ theorem~\cite{wilksLargeSampleDistributionLikelihood1938b}. This enables selection of the \emph{minimal adequate} model: the simplest model whose likelihood is not significantly improved by adding parameters associated with additional physical mechanisms, such as correlated dissipation or higher-order interactions.

To make this procedure computationally practical, we employ end-to-end differentiation through an ODE-based simulator of the experiment. This allows the likelihood to be optimized directly through the simulated experiment. In practice, gradients are computed through state preparation, Lindblad evolution, and measurement, and are used to fit $k$-local structured Hamiltonians and dissipators with automatic differentiation and adjoint methods \cite{bradburyJAXComposableTransformations2018, kidgerNeuralDifferentialEquations2022b}. As a result, candidate models can be fitted and compared within the same differentiable workflow, even as the hierarchy is enlarged to include higher-locality Hamiltonian terms, correlated dissipation, or latent degrees of freedom.

We demonstrate the LIMINAL framework on a superconducting five-qubit device arranged in a weight-4 parity layout in three settings. First, using time-resolved tomographic data for idling dynamics, we identify the minimal adequate k-local Lindblad model for simultaneously describing  all five qubits. Second, we validate the likelihood-based inference in well-understood driven single-qubit experiments. We reconstruct both time-independent generators under square pulses as well as a time-dependent Hamiltonian of a DRAG-corrected pulse with a cosine envelope. Third, we extend the approach to qubit-coupler dynamics influenced by hidden subsystems by enlarging the Hilbert space with latent two-level systems; LR comparisons then test whether hidden degrees of freedom and higher interaction locality are required to explain coupler-mediated entangling interactions.

\begin{figure}
    \centering
    \includegraphics[width = 1.0\linewidth]{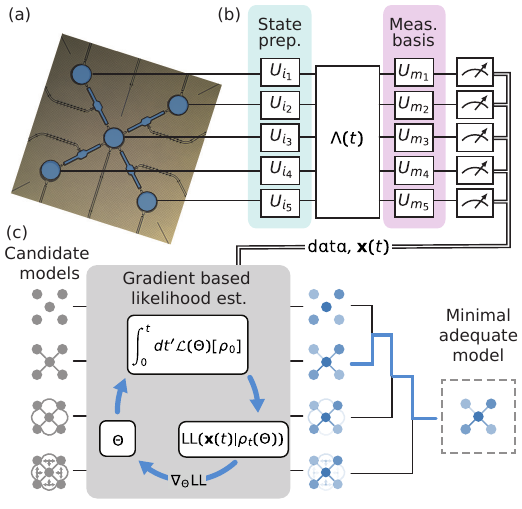}
    \caption{Conceptual overview of the framework for learning the minimal Lindblad model.
    \textbf{(a)} Optical micrograph of the superconducting 5-qubit device with fixed-frequency transmons and tunable couplers used in this work. Further experimental details can be found in Supplemental Material \ref{sec:setup}.
    \textbf{(b)} Tomographic measurement sequence: Qubits are prepared using local preparation gates $U_i$, and the system is allowed to evolve under the time-dependent channel of interest, $\Lambda(t)$. Finally, local basis-change gates $U_m$ are applied to each qubit before  projective readout is performed.
    \textbf{(c)} Classical post-processing of the data. A hierarchy of nested candidate Lindblad models (gray pictograms) are suggested. Each model is fitted using a gradient-based maximum-likelihood estimation of the observed data enabled by a differentiable ODE simulator. Models with fitted parameterizations (blue pictograms) are tested against each other using likelihood ratio tests. When no additional explanatory power is observed, the smallest model that is adequate for describing the dynamics of the quantum processor is assumed.}
    \label{fig:concept}
\end{figure}

\section{Results}
\subsection{Conceptual Description of LIMINAL}\label{sec:concept}
This section introduces LIMINAL as a framework for identifying the simplest model that accurately describes the time-dependent dynamics of a quantum processor. We exemplify the framework using the 5-qubit processor shown in Fig.~\ref{fig:concept}a, which is used for the experimental demonstrations in sections~\ref{sec:5qubit_idle}--\ref{sec:hidden_qubits}. The experiment probes a dynamical map $\Lambda(t)$ that takes a prepared input state to the state reached after an evolution time $t$. In our models, this map is described by evolving the state according to a Lindblad master equation \cite{lindbladGeneratorsQuantumDynamical1976a},
\begin{equation}\label{eq:lindblad}
\frac{d\rho}{dt}
=\mathcal{L}(\rho)=
-\frac{i}{\hbar}[H,\rho]
+
\mathcal{D}(\rho),
\end{equation}
where $\mathcal{L}$ is the Lindblad generator, \(H\) denotes the Hamiltonian contribution and \(\mathcal{D}\) collects dissipative terms. LIMINAL tests which parameterized descriptions of these contributions are warranted by the data. The procedure has three parts: (i) define a hierarchy of nested candidate models, (ii) collect time-resolved tomographic data sensitive to the parameters of the models, and (iii) compare their explanatory power using likelihood-ratio (LR) tests. 

To ensure sensitivity to any candidate model within the considered hierarchy, we acquire a tomographically complete dataset for the dynamics of interest (Fig.~\ref{fig:concept}b) by repeating the following sequence:
\begin{enumerate}
    \item Reset all $N$ qubits to the ground state.
    \item Prepare each qubit by applying a product of single-qubit gates, $\otimes_n^N U_{i_n}$, spanning an informationally complete basis of the $N$-qubit system (e.g., cardinal states along $x$, $y$, and $z$).
    \item Let the dynamics of interest evolve for a time interval $t$, defining the experimentally probed channel $\Lambda(t)$.    
    \item Measure each qubit using an informationally complete POVM, by applying a basis‑change gate, $\bigotimes_{n}^N U_{m_n}$, followed by projective readout of $z$.
    \item Record the resulting bit string, $b$, for classical analysis.
\end{enumerate}

This procedure is repeated over all configurations of initial states, measurement bases, and time intervals (from $t=0$ to the full process duration with suitably chosen discretization). Additionally, each configuration is repeated $N_\text{shots}$ times to obtain statistics. The outcomes are stored in a vector $\mathbf{x}_{i, m}(t)\in\mathbb{N}^{2^N}$, whose entries are the number of times a particular bitstring, $b \in \{0, 1\}^N$ occurs. Because we choose informationally complete initialization and measurement settings, the number of configurations scales exponentially with system size \cite{parisQuantumStateEstimation2004, harrowLimitationsQuantumDimensionality2015}.

We now describe the classical analysis illustrated in Fig.~\ref{fig:concept}c. A \emph{parameterization}, denoted \(\theta\), is a structured set of real parameters specifying one physical object, such as an initial state, Hamiltonian, or dissipator. We write \(\Theta\) for the collection of parameterizations used in a candidate description of the experiment, \(\Theta=\{\theta_\rho,\theta_H,\theta_L\}\), where \(\theta_\rho\) contains parameters related to the state, \(\theta_H\) contains Hamiltonian parameters, and \(\theta_L\) contains dissipation-related parameters. A \emph{model} is then the differentiable map from \(\Theta\) to simulated density matrices, predicted measurement probabilities, and hence a likelihood value. The chosen hierarchy of \emph{nested} models (grey pictograms) should satisfy that each successive model extends the previous one by adding parameters, so that the associated parameter spaces \(\mathcal S\) satisfy \(\mathcal{S}_A \subset \mathcal{S}_B\) for models \(A\) and \(B\).

Each candidate model is fitted by gradient-based maximum-likelihood estimation. Given an initial choice of \(\Theta\), the model predicts, for each configuration of preparation \(i\), measurement basis \(m\), and time \(t\), a density matrix \(\rho_{i,m}(t;\Theta)\). This in turn defines a probability vector, \(\mathbf{p}_{i,m}(t;\Theta)\), via Born's rule over the \(2^N\) bit-string outcomes. A Gaussian approximation to the count statistics is reliable only when the relevant expected bin counts \(N_{\mathrm{shots}}p_b\) are sufficiently large. For an experiment with \(2^N\) possible bit-string outcomes, this condition can require many repetitions whenever the probability mass is broadly distributed. We therefore model the observed count vector directly as multinomially distributed with probabilities \(\mathbf{p}_{i,m}(t;\Theta)\), giving the log-likelihood
\begin{align}
\mathrm{LL}(\Theta)
=
\sum_{i,m, t}
\log \mathrm{Multinomial}\!\left(
\mathbf{x}_{i,m}(t)\,;\,
\mathbf{p}_{i,m}(t;\Theta),\,
N_{\mathrm{shots}}
\right).
\end{align}
The gradient of \(\mathrm{LL}(\Theta)\) with respect to \(\Theta\) is obtained by differentiating through the entire model. Parameters are iteratively updated using gradients information till convergence, ultimately yielding the \emph{maximum-likelihood estimate} (MLE) of the parameters. By inverting the Hessian matrix evaluated at the estimated parameters, error estimates can be extracted \cite{efronAssessingAccuracyMaximum1978}. Further details are given in Section~\ref{sec:gradient}. 

After fitting, we compare the log-likelihood values of nested models while accounting for differences in complexity using LRs and Wilks' theorem~\cite{wilksLargeSampleDistributionLikelihood1938b}. Alternative model-selection criteria such as Akaike information criterion (AIC) and Bayesian information criterion (BIC) penalize complexity without requiring a nested structure~\cite{akaikeNewLookStatistical1974, schwarzEstimatingDimensionModel1978}. Here we use LR comparisons because our candidate families are explicitly nested and LR tests provide a direct likelihood-based way of quantifying if additional parameters are warranted by the data. Under Wilks' theorem, if the smaller model is correct, twice the increase in log-likelihood achieved by the larger model,
\begin{equation}
2\Delta\mathrm{LL} = 2\big(\mathrm{LL}_{\text{large}} - \mathrm{LL}_{\text{small}}\big),
\end{equation}
is asymptotically distributed as a chi-square distribution, \(\chi^2_{\Delta d}\), where \(\Delta d\) is the number of additional \emph{independent real degrees of freedom}, in the larger model compared to the smaller model. However, in practice, formal \(p\)-values based on direct comparison with \(\chi^2_{\Delta d}\) can be sensitive to deviations from this ideal asymptotic chi-square approximation, including finite-sampling, numerical artifacts, and model-regularity effects. We therefore do not report \(p\)-values and instead quantify improvement via the \emph{explanatory power}~\cite{hashimPracticalIntroductionBenchmarking2025a}
\begin{equation} \label{eq:EP}
\Xi \;=\; \frac{2\Delta\mathrm{LL} - \Delta d}{\sqrt{2\Delta d}}\,,
\end{equation}
which measures the improvement of a model extension relative to the Wilks reference value and scale. Under the ideal Wilks asymptotics, the explanatory power will approximate a normal distribution with unit variance \(\Xi \approx \mathcal{N}(0,1)\) for large \(\Delta d\). Evaluating \(\Xi\) across successive model extensions therefore identifies the first extension that provides no meaningful explanatory power and, by extension, the simplest adequate model for the dataset. Choosing a decision threshold necessarily involves a convention rather than a theorem-implied cutoff; as a practical guideline, a 5\% one-sided criterion in the Gaussian approximation corresponds to \(\Xi_{\mathrm{th}} \approx 1.65\), which we use to indicate when an extension provides non-negligible explanatory power.

LIMINAL thus provides a likelihood-based framework for testing competing dynamical descriptions of quantum-processor behavior. In this work, however, we restrict attention to processor dynamics under the following assumptions: (i) the dynamics remain within the computational subspace of dimension $2^N$, (ii) state preparation and measurement (SPAM) errors are modeled with imperfect state preparation but ideal measurements, and (iii) single-qubit unitary operations are modeled as ideal (see benchmarking in supplemental material \ref{sec:characterization-benchmarking}). These assumptions can be relaxed and tested within the LIMINAL framework, but we maintain them here to limit scope. Additionally, we consider the dynamics in a rotating frame of the drive frequencies of the single-qubit control gates.

\subsection{Model Testing for an Idling 5 Qubit Processor}\label{sec:5qubit_idle}
\begin{figure*}
    \centering
    \includegraphics[width=1.0\linewidth]{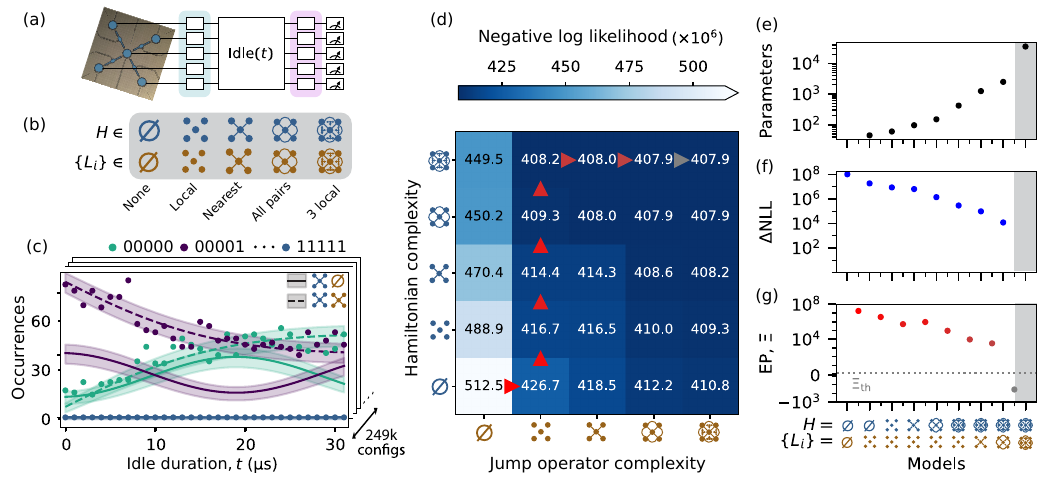}
    \caption{
    Application of the LIMINAL framework to determine the minimal adequate model for the idling dynamics of a 5-qubit processor. 
    \textbf{(a)} Tomographic experiment for idling dynamics on the 5‑qubit device.
    \textbf{(b)} Candidate models arranged along two axes: unitary and dissipative locality. Complexity increases from local terms to nearest-neighbor (two-local), all-to-all two-local, and three-local interactions. 
    \textbf{(c)} Observed bit-strings from one example configuration and fitted predictions from two different models.   
    \textbf{(d)} Log-likelihood values for all model combinations, where complexity increases from lower left to upper right. Arrows indicate the path of greatest explanatory power. See main text for details.
    \textbf{(e)} Number of parameters, \textbf{(f)} difference in negative log-likelihood to the lowest values achieved and \textbf{(g)} explanatory power for model extensions along this path. The horizontal dashed line marks the threshold $\Xi_\mathrm{th}$ for accepting a model extension as significant, while the gray region indicates extensions that do not provide significant explanatory power.
    }
    \label{fig:model_selection}
\end{figure*}

In this section, we apply the LIMINAL framework to a 5-qubit processor (Fig.\ \ref{fig:concept}a) to identify the simplest Lindblad model that sufficiently captures its idling dynamics. Our objectives are twofold: first, to demonstrate that gradient-based likelihood estimation enables fitting Lindblad dynamics at this scale, and second, to identify the point at which increasing model complexity of a 5-qubit processor ceases to provide explanatory power.

We model the idling dynamics by a time-independent Markovian Lindblad generator. 
To obtain a concrete parameterization, we expand the generator in the Pauli-string basis over all five qubits. 
We write \(P_i = \bigotimes_{n=1}^N \sigma_{i n}\), where \(\sigma_{i n}\in\{I,X,Y,Z\}\), and exclude the all-identity string from the indexed set. 
Following Ref.~\cite{samachLindbladTomographySuperconducting2022a}, this yields
\begin{align}\label{eq:lindblad_pauli}
    \frac{d\rho}{dt}
    =
    -\frac{i}{\hbar}\sum_i a_i [P_i, \rho]
    +
    \sum_{m,n} D_{mn}
    \left(
        P_m \rho P_n
        - \frac{1}{2}\{P_n P_m,\rho\}
    \right).
\end{align}
Here \(\rho\) is the five-qubit density matrix, \(\{P_i\}\) denotes the non-identity Pauli strings, \(a_i\) are real Hamiltonian coefficients, and \(D_{mn}\) is a positive semidefinite Hermitian matrix encoding dissipation. 
By diagonalizing \(D\), the dissipative term can equivalently be written in the usual jump-operator form,
\begin{equation}
\mathcal{D}(\rho)
=
\sum_i \Gamma_i
\left(
L_i \rho L_i^\dagger
-
\frac{1}{2}\{L_i^\dagger L_i,\rho\}
\right),
\end{equation}
with non-negative rates \(\Gamma_i\) and the set of jump operators $\{L_i\}$. 
This representation fully specifies the Lindblad generator: the Hamiltonian requires \(4^N-1\) parameters, while the dissipation matrix involves \((4^N-1)^2\) parameters. 
For five qubits, this corresponds to 1,047,552 parameters, excluding SPAM parameters, rendering an unconstrained fit impractical.

To address this challenge, we impose structured parameterizations based on locality (the number of non-identity Pauli operators in the strings $\{P_i\}$) and the qubit connectivity graph. The Hamiltonian is written as a sum of sub-Hamiltonians grouped by locality:
\begin{equation}\label{eq:ham_param}
H = \sum_k^{k_\text{max}} H^{(k)}
\end{equation}
where $H^{(k)}$ collects all $k$-local terms, i.e.\ terms with exactly $k$ non-identity Pauli operators. Each locality sector can further be restricted by a graph structure, such as nearest-neighbor or all-to-all connectivity. For each locality $k$, the corresponding coefficients are collected in a $(1+k)$-dimensional array $a^{(k)}$, where the first axis indexes the allowed $k$-qubit connections in the graph and the remaining $k$ axes index the Pauli components on those qubits. The full Hamiltonian parameterization is therefore the collection
\[
\theta_H = \{a^{(1)}, a^{(2)}, \dots, a^{(k_\text{max})}\},
\]
describing one-local, two-local, and higher-order interactions. This locality-based structure reduces the parameter space from exponential to polynomial scaling, $O(N^{k_\text{max}})$, provided the maximum locality $k_\text{max}$ remains fixed~\cite{baireyLearningLocalHamiltonian2019}. Further details of the Hamiltonian parameterization are given in Section~\ref{sec:param}.

We adopt an analogous locality‑based construction for dissipation, which can be chosen independently of the Hamiltonian. For a chosen maximum locality $k_\text{max}$, we define blocks of jump operators acting on $k_\text{max}$-qubit subsystems according to the connection graph, $\mathcal{C}_{k_\text{max}}$. This ensures that the parameterization includes all dissipative effects of locality $k$ up to $k_\text{max}$-locality, since lower-locality terms are naturally contained within these blocks. Formally, the full set of jump operators for locality $k_\text{max}$ is
\begin{equation}\label{eq:jump_param}
\{ L_i^{(k_\text{max})} \} = \bigcup_{c \in \mathcal{C}_{k_\text{max}}} \{ L_i^{(c)} \},
\end{equation}
where \(\mathcal{C}_{k_\text{max}}\) is the set of all allowed \(k_{k_\text{max}}\)-qubit connections in the interaction graph, and each block $\{ L_i^{(c)} \}$ corresponds to the jump operators generated for $k_{k_\text{max}}$-local subset, $c$. This construction provides a nested hierarchy of dissipative models analogous to the Hamiltonian case. Further details on the parameterization and its numerical implementation are provided in Section~\ref{sec:param}. 

In Fig.~\ref{fig:model_selection}a, we probe the simultaneous 5-qubit idling channel using the tomographic protocol of Sec.~\ref{sec:concept}. To reduce the memory footprint, we use four single-qubit initial states instead of the overcomplete set of Pauli eigenstates. Since any set of preparations spanning the Bloch sphere is sufficient, we choose the vertices of a regular tetrahedron inscribed in the Bloch sphere, i.e.\ single-qubit SIC states, which treat all directions symmetrically (see Supplemental Material \ref{sec:characterization-benchmarking} and Ref.~\cite{renesSymmetricInformationallyComplete2004}). This yields 248,832 configurations. For each configuration, the idling dynamics are measured from \(t=0\) to \(t=\SI{30}{\micro s}\) in \(\SI{1}{\micro s}\) steps with \(N_{\mathrm{shots}}=100\) repetitions.

The candidate models for describing the observed data are shown in Fig.~\ref{fig:model_selection}b. The models are arranged along two axes: unitary (blue) and dissipative (gold) with k-local graphs spanning none, single-qubit interactions, nearest-neighbor, all-to-all and all three-body interactions. As prescribed in Section~\ref{sec:gradient}, we perform gradient‑based MLE through the differentiable ODE simulator for each combination. To reduce memory use during optimization, we compute each gradient update using a random minibatch containing one sixteenth of the initial-state preparations, while the final log-likelihood of each fitted model is evaluated on the full dataset.

Fig.~\ref{fig:model_selection}c shows observed data for one configuration alongside predictions from two fitted models: (1) a nearest-neighbor Hamiltonian with no dissipation (solid line), and (2) a nearest-neighbor Hamiltonian with nearest-neighbor dissipation (dashed line). The lines illustrate the median predicted number of occurrences and the shaded region the 68\% confidence region extracted from the simulation. In the no dissipation/local Hamiltonian model (solid), the fit fails to finds a descriptive set of parameters capturing the exponential decay. Adding the dissipation allows for a much better description of the data (dashed line). In addition to this configuration, we provide 18 additional, randomly selected configurations in the supplemental material \ref{sec:idling_supp}.

We now expand our models to find the minimal adequate model. Fig.~\ref{fig:model_selection}d reports the negative log‑likelihood by applying the likelihood estimation to each model combination. The reported values are the negative log-likelihood evaluated on the entire dataset. The lower left corner consist of a model with no dissipation and no interactions. The upper right corner has all three local interaction and three local dissipation. By moving from the simplest model to the largest we apply the model testing framework to stop at the simplest adequate description of our data. Starting from the lower left corner, the arrows indicate the direction that maximizes the explanatory power, $\Xi$ (Eq.\ \eqref{eq:EP}). The path now prioritizes the possible additions to our model and provides a stopping point when no explanatory power is found by extending the description further.
The largest improvement is obtained by including \emph{local dissipation} channels. Physically, this corresponds to the dominant single-qubit error mechanisms, such as relaxation and dephasing, that are required to explain the observed decay of populations and coherences during idling. The next improvements arise from adding \emph{local Hamiltonian} terms, consistent with static single-qubit frequency offsets and control-frame miscalibration, followed by \emph{nearest-neighbor Hamiltonian} interactions, consistent with residual couplings on the device graph, and then by extending the two-local Hamiltonian terms to \emph{all-pairs} connectivity.

At this stage, adding nearest-neighbor dissipation reduces the negative log-likelihood the most, but explanatory power is maximized instead by extending the Hamiltonian to \emph{three-local} terms, since this yields a larger gain relative to the smaller parameter increase. Along the selected path, the next accepted dissipative extensions are then nearest-neighbor dissipation followed by all-to-all two-local dissipation. Beyond these extensions, the explanatory power vanishes: adding \emph{three-local dissipation} provides no meaningful improvement. This saturation is directly visible as \(\Xi \le \Xi_\mathrm{th}\) for the corresponding extensions indicating that the additional dissipative parameters are not warranted by the data.

Fig.~\ref{fig:model_selection}e--g summarizes the trade-off between model complexity and fit quality along the selected model path. Fig.~\ref{fig:model_selection}e shows the number of free parameters in each successive model, which increases rapidly as additional Hamiltonian and dissipative terms are included. Fig.~\ref{fig:model_selection}f reports the difference in negative log-likelihood between each model and the best value achieved; this gap decreases as the models become more expressive, but the improvement eventually saturates, indicating diminishing returns from further extensions. This is made explicit in Fig.~\ref{fig:model_selection}g, which shows the explanatory power, $\Xi$, associated with each successive model extension. The largest values of $\Xi$ occur for the early extensions, showing that the dominant gains arise from introducing local dissipation and progressively richer Hamiltonian structure. Once three-local Hamiltonian terms are included, the remaining explanatory gains come from increasing the complexity of the dissipative sector. Specifically, two-local dissipation provides explanatory power for both nearest-neighbor and all-to-all connectivity, whereas three-local dissipation does not. Together, these three panels show that beyond the selected model with three-local Hamiltonians and two-local dissipation on an all-to-all connectivity graph, the additional parameters increase complexity more than they improve the description of the data. We therefore conclude that, within the tested model family, the idling dynamics of our superconducting QPU are adequately captured by a Lindblad model with three-local Hamiltonians and two-local dissipation, while three-local dissipation is not warranted by the data.

In this example, we acquired a tomographically complete dataset and fitted every model in the hierarchy. 
We stress, however, that LIMINAL can be implemented adaptively: after fitting a baseline model, one designs the next round of measurements to maximize sensitivity to the additional parameters in the proposed extension (while remaining informative for the already-estimated ones). By focusing data collection on the degrees of freedom that differentiate competing models, one can substantially reduce both the required experimental configurations and the overall fitting cost.

\subsection{Reconstruction of a Constant Pulse}\label{sec:const}
\begin{figure}
    \centering    \includegraphics{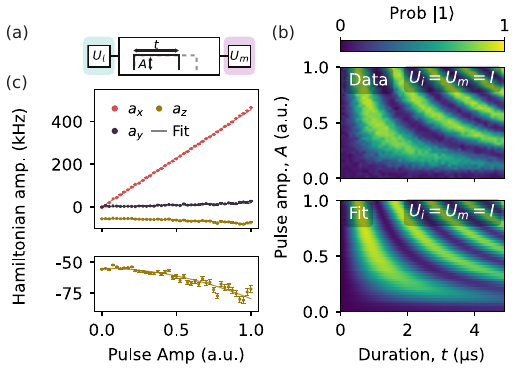}
    \caption{
    Single-qubit validation under a constant, slightly detuned pulse.
    \textbf{(a)} Tomographic sequence: local preparation $U_i$, square pulse of amplitude $A$ and duration $t$, basis-change $U_m$, and projective readout.
    \textbf{(b)} Probability of measuring the excited state as a function of time for different pulse amplitudes. Experimental data (top) are compared with simulated fits (bottom). 
    \textbf{(c)} Fitted Hamiltonian parameters for each amplitude, decomposed into Pauli components. A zoom-in highlights the $z$-component (gold), while the $x$- and $y$-components (red and blue) are fitted with a linear model and the $z$-component with an offset quadratic form.
    }
    \label{fig:const}
\end{figure}

With the idling dynamics characterized, we next model the processor's response to applied control fields. As a validation step, we reconstruct the well-understood response of a single qubit under a constant, slightly detuned drive~\cite{krantzQuantumEngineersGuide2019b, chiorescuCoherentQuantumDynamics2003}. We test whether the likelihood-based inference procedure of LIMINAL can infer physically meaningful Hamiltonian components (and dissipation) from time-resolved tomographic data before moving to time-dependent and hidden-subsystem scenarios in Secs.~\ref{sec:gates} and~\ref{sec:hidden_qubits}, respectively.

Fig.~\ref{fig:const}a sketches the experiment. A single qubit is driven by a square pulse at a drive frequency $100\,\mathrm{kHz}$ from the qubit transition. In the rotating frame of the drive, this yields oscillations predominantly about the Pauli-$x$ direction together with a fixed frequency offset, $\delta f = f_{\mathrm{frame}} - f_{\mathrm{qubit}} = 100\,\mathrm{kHz}$. We repeat the experiment for multiple pulse amplitudes $A$, thereby increasing the oscillation rate while keeping the detuning fixed. For each amplitude, we collect a tomographically complete dataset by combining informationally complete state preparations and measurement bases (Section~\ref{sec:concept}) over a range of pulse durations.

In Fig.~\ref{fig:const}b, we show the dynamics for one representative tomographic configuration, $U_i = U_m = \mathbb{I}$, as a function of pulse duration and pulse amplitude. The upper panel shows the measured excited-state probability, while the lower panel shows the corresponding dynamics predicted by the fitted model. As the pulse amplitude increases, the oscillation frequency increases, as expected for driven Rabi oscillations measured in the $z$-basis. For each amplitude $A$, the fit is obtained by maximum-likelihood estimation within the full single-qubit, time-independent Lindblad model of Eq.~\eqref{eq:lindblad_pauli}, using gradient-based optimization and warm-starting from the previous amplitude. The agreement between the measured and fitted traces shows that this fixed model family captures the amplitude dependence of the driven dynamics well. All 18 tomographic configurations are shown in the supplemental material~\ref{sec:const_supp}.

Fig.~\ref{fig:const}c shows the fitted Hamiltonian coefficients in the Pauli basis as a function of pulse amplitude $A$ (points), together with simple trend lines over the scanned amplitude range. The coefficients $a_x$ (red) and $a_y$ (blue) are fitted with linear functions, while $a_z$ (gold) is fitted with a quadratic form; the lower panel provides a zoom of $a_z$ for clarity. The reconstructed coefficients follow the expected behavior, $a_x$ increases approximately linearly with amplitude, consistent with the applied transverse drive, while $a_y$ remains zero within the resolution of the reconstruction. In contrast, $a_z$ exhibits a quadratic dependence on $A$, consistent with an AC Stark shift arising from weak anharmonicity and coupling to higher levels in superconducting qubits~\cite{motzoiSimplePulsesElimination2009a}. At zero drive amplitude, the offset $2a_z(A=0)\approx 100\,\mathrm{kHz}$ matches the detuning of the rotating frame relative to the qubit frequency. 

The experiments performed in this section confirm that the framework recovers physically interpretable parameters under driven conditions. This establishes confidence in the framework before moving to time-dependent Hamiltonian reconstruction in Section~\ref{sec:gates} and hidden-subsystem analysis in Section~\ref{sec:hidden_qubits}.

\subsection{Reconstructing a Time Dependent Envelope}\label{sec:gates}

\begin{figure}
    \centering    \includegraphics{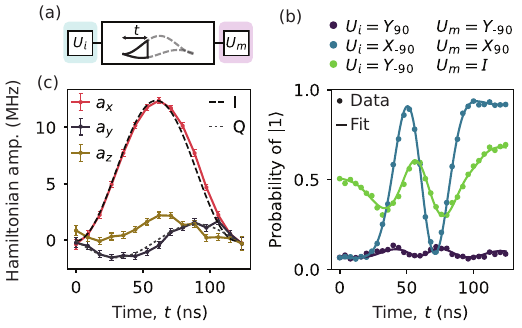}
    \caption{Reconstruction of a time-dependent Hamiltonian from an enveloped resonant drive.
    \textbf{(a)} Tomographic sequence: local preparation $U_i$, resonant pulse with a shaped $\text{I}(t)$ envelope; the $\text{Q}(t)$ quadrature follows a DRAG-like derivative scaling. The pulse is truncated at time $t$ in fixed increments to map the evolving dynamics, followed by basis-change $U_m$ and projective readout.
    \textbf{(b)} Representative configurations showing data (points) and model predictions (solid lines) obtained by maximum likelihood estimation.
    \textbf{(c)} Reconstructed single-qubit Hamiltonian components $a_x(t)$, $a_y(t)$, $a_z(t)$, obtained via piecewise-linear interpolation between anchor points. The I/Q envelopes are overlaid (scaled) for comparison.
    }
    \label{fig:1qubit_menu}
\end{figure}

Realistic control pulses for quantum gates are typically shaped to suppress leakage and improve gate fidelity \cite{hyyppaReducingLeakageSingleQubit2024, motzoiSimplePulsesElimination2009a, gambettaAnalyticControlMethods2011}. Conventionally, such pulse engineering is assessed indirectly from the net effect of the fully applied gate on the qubit, which reveals the overall action of the pulse. Here, instead, we truncate the pulse at intermediate times to reconstruct the driven Hamiltonian \emph{during} the application. Building on the time-independent validation in Section~\ref{sec:const}, we show that our likelihood-based inference can recover a time-dependent Hamiltonian directly from tomographic data, without imposing an explicit analytic model for the pulse shape. This provides a route to characterizing the effective control fields experienced by the qubit for experimentally relevant shaped pulses.  

Fig.~\ref{fig:1qubit_menu}a illustrates the experiment; we apply a resonant drive in the in-phase (I) quadrature with a cosine envelope. The quadrature (Q) component is chosen proportional to the time-derivative of the I-envelope, following the DRAG prescription~\cite{motzoiSimplePulsesElimination2009a}. To obtain time-resolved information about the evolving dynamics, we truncate the pulse at time $t$ in $4\,\mathrm{ns}$ increments up to a total duration of $120\,\mathrm{ns}$. For each truncation time, we collect a tomographically complete dataset and repeat the full dataset $N_\text{shots}=1000$ times. 

In Fig.~\ref{fig:1qubit_menu}b, we show representative tomographic configurations for the shaped-pulse experiment, with measured data shown as points and fitted dynamics shown as solid lines. The pulse truncation time on the horizontal axis makes the evolving qubit response directly visible, while different preparation and measurement settings provide sensitivity to different components of the driven dynamics. The agreement between data and fit across these representative traces shows that the inferred model captures the time dependence of the control field. All tomographic configurations are shown in the supplemental material~\ref{sec:gate_supp}.

In Fig.~\ref{fig:1qubit_menu}c, we show the reconstructed time-dependent Hamiltonian in the Pauli basis, with coefficients $a_x(t)$, $a_y(t)$, and $a_z(t)$ plotted as functions of time with the I and Q quadrature of the intended pulse envelope shown with dashed lines for comparison. To model this time dependence, we extend the single-qubit Hamiltonian of Sec.~\ref{sec:const} to allow explicitly time-dependent coefficients,
\begin{equation}
H(t) = a_x(t)\,\sigma_x + a_y(t)\,\sigma_y + a_z(t)\,\sigma_z.
\end{equation}
To remain agnostic to the pulse shape, each coefficient is parameterized by values at 15 equally spaced anchor points in time, shown as markers in Fig.~\ref{fig:1qubit_menu}c. Between anchor points, the Hamiltonian is linearly interpolated, shown as solid lines, yielding a continuous time-dependent model with a total of 45 Hamiltonian parameters.  Because LIMINAL already uses an ODE solver for the forward simulation, incorporating $H(t)$ requires no additional modifications beyond specifying the interpolation parameterization. This parameterization allows the qubit itself to act as a probe of the effective control fields experienced during the shaped pulse.

The reconstructed coefficients show that $a_x(t)$ and $a_y(t)$ track the intended envelopes in the I and Q quadratures, respectively, while $a_z(t)$ is non-negligible and follows the magnitude of the pulse envelope qualitatively. As in Sec.~\ref{sec:const}, we interpret this longitudinal component as an AC Stark shift. Further, we observe a small delay on the falling edge of the reconstructed Hamiltonian relative to the nominal pulse timing, this is attributed to transient ringing induced by pulse truncation. 

Overall, this experiment shows that our likelihood-based inference method can reconstruct an effective time-dependent single-qubit Hamiltonian directly from tomographic data without imposing an analytic pulse model. Beyond validating the intended control envelope, this provides a route to diagnosing pulse distortions and characterizing the control fields actually experienced by the qubit during time-dependent operations.

\subsection{Studying Hidden-Qubit Models in Coupler Mediated Entangling Dynamics}\label{sec:hidden_qubits}
\begin{figure*}
    \centering
    \includegraphics[width=1.0\linewidth]{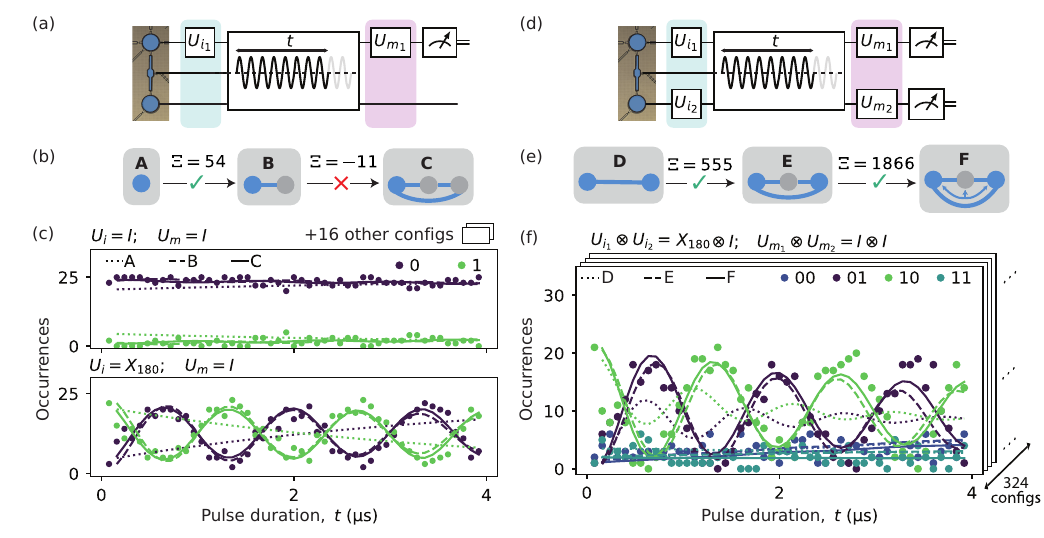}
    \caption{
    Capturing coupler-mediated entangling dynamics and non-Markovian memory with minimal hidden qubit models.
    \textbf{(a)} Two-qubit experiment where only the properties of qubit 1 is tomographically reconstructed. Two fixed-frequency transmons are connected by a tunable coupler. A flux modulation at the qubit-frequency difference induces an iSWAP-like interaction. Tomography is performed on the first qubit only.
    \textbf{(b)} Nested models for one observed qubit. Model \emph{A}: single-qubit Lindblad model. Model \emph{B}: the observed qubit coupled to one hidden two-level system (TLS). Model \emph{C}: the observed qubit coupled to two hidden TLSs.
    \textbf{(c)} Observed data for two configurations alongside model-predicted probabilities for the estimated models \emph{A}–\emph{C}.
    \textbf{(d)} Same experiment but now with tomographic measurements of two qubits.
    \textbf{(e)} Nested models for two observed qubits. Model \emph{D}: effective two-qubit model. Model \emph{E}: three-qubit model with one hidden qubit coupled to both observed qubits. Model \emph{F}: three-local Hamiltonian interaction allowed in the three-qubit model.
    \textbf{(f)} Observations and predictions for a configuration where the first qubit is initialized with an $X_\pi$  gate.}
    \label{fig:coupler}
\end{figure*}

Additional degrees of freedom outside the observed qubit subspace—ranging from parasitic two-level systems to engineered circuit elements such as resonators and tunable couplers—can strongly influence the measured processor dynamics \cite{siddiqiEngineeringHighcoherenceSuperconducting2021, mckayUniversalGateFixedFrequency2016a, blaisCircuitQuantumElectrodynamics2021b}. When such degrees of freedom can be treated as a Markovian environment, their effect can be absorbed into the Lindblad master equation (Eq.\ \eqref{eq:lindblad_pauli}). However, when they retain memory on experimental timescales, the reduced dynamics of the observed qubits become non-Markovian and a Lindblad model on the observed Hilbert space is insufficient to describe the system \cite{odehNonMarkovianDynamicsSuperconducting2025}. Non-Markovian extensions include explicit memory kernels~\cite{nakajimaQuantumTheoryTransport1958,zwanzigEnsembleMethodTheory1960} or time-local master equations~\cite{varonaLindbladlikeQuantumTomography2025b}. Here we pursue an alternative that remains compatible with our ODE-based likelihood framework: we enlarge the Hilbert space with \emph{hidden} qubits (latent two-level systems) and trace them out before likelihood evaluation \cite{shnirmanLowHighFrequencyNoise2005, anto-sztrikacsCapturingNonMarkovianDynamics2021}. With LIMINAL, this allows us to test, whether hidden subsystems are required and what k-local interaction is warranted by the data.

In Fig.~\ref{fig:coupler}, we consider a coupler-mediated interaction experiment in which latent degrees of freedom are expected to matter. Two fixed-frequency transmons are connected by a tunable coupler, and a flux modulation is applied to the coupler at the qubit-frequency difference, inducing an interaction analogous to an iSWAP gate~\cite{mckayUniversalGateFixedFrequency2016a, warrenExtensiveCharacterizationImplementation2023a}. We analyze two datasets acquired from this same physical setting: (i) \emph{single-qubit tomography} (Fig.~\ref{fig:coupler}a-c), where measurements are performed only on one qubit while the second transmon remains unobserved, and (ii) \emph{two-qubit tomography} (Fig.~\ref{fig:coupler}d-f), where both qubits are measured tomographically. In both cases, the central question is whether the observed statistics can be explained by an effective Markovian model on the observed subsystem, or whether hidden degrees of freedom must be included.

Although the underlying interaction is time-dependent, we seek an \emph{effective time-independent} description that captures the measured dynamics over the experiment duration without invoking additional rotating-frames or perturbative reductions. Conventional treatments often simplify coupler-mediated interactions using rotating-frame and Schrieffer-Wolff transformations \cite{schriefferRelationAndersonKondo1966, rothAnalysisParametricallyDriven2017}. Here we instead use LIMINAL to learn an effective generator directly from measurement data and use LR comparisons to assess what model structure is warranted. 
All models including at least two-qubit interactions also incorporate two-qubit correlated dissipation, consistent with the findings in Section~\ref{sec:5qubit_idle}.

In Fig.~\ref{fig:coupler}b, we compare three nested models for the single-qubit tomography scenario based on the time-independent Lindblad master equation, Eq.~\eqref{eq:lindblad_pauli}:
\begin{itemize}
    \item \textbf{Model A}: single-qubit model.
    \item \textbf{Model B}: two-qubit model including one hidden qubit.
    \item \textbf{Model C}: three-qubit model including two hidden qubits, with all three qubits interacting through 2-local terms.
\end{itemize}

Fig.~\ref{fig:coupler}c shows time traces for two initial-state configurations: $U_i=\mathbb{I}$ and $U_i={X_\pi}$, both measured in the $z$-basis ($U_m=I$). Model A reproduces the dynamics of the $\ket{0}$ state (purple) but does not capture the oscillation when initialized in $\ket{1}$ (green). Model B resolves these inconsistencies by introducing a hidden qubit that captures memory effects, while Model C adds complexity without improving likelihood, as confirmed by the explanatory power metric $\Xi=-11<\Xi_\mathrm{th}$. The complete tomographic data is available in the supplemental material \ref{sec:coupler_supp}. 

In Fig.~\ref{fig:coupler}e, we compare three nested models for the full two-qubit tomographic dataset. In the standard effective description of a coupler-mediated gate, the coupler degree of freedom is eliminated, yielding an effective two-qubit model for the observed transmons. Here, we test whether this reduced description is sufficient to explain the data, or whether the coupler must be reintroduced as a hidden degree of freedom to capture the dynamics more faithfully.
\begin{itemize}
    \item \textbf{Model D}: effective two-qubit model.
    \item \textbf{Model E}: three-qubit model with pairwise couplings, where the coupler qubit is hidden.
    \item \textbf{Model F}: three-qubit model with a hidden coupler qubit and a global interaction graph allowing 3-local Hamiltonian terms.
\end{itemize}

In Fig.~\ref{fig:coupler}f, we show representative time traces for initialization in the state $\ket{10}$ and measurement in the $z$-basis. Model D provides a baseline fit but fails to reproduce the observed oscillations. Model E yields a substantial improvement, showing that an effective two-qubit description is not sufficient and that hidden coupler dynamics is required to account for the data. Extending the model further to Model F produces an additional increase in explanatory power, indicating that 3-local Hamiltonian terms contribute meaningfully to the observed dynamics. This shows that, within the tested model family, both the hidden coupler degree of freedom and higher-order interaction terms improve the description of the experiment. However, because the explanatory power remains significant at the largest model considered, additional model extensions would be needed to determine whether Model F is already adequate. Additional tomographic configurations are shown in the supplemental material~\ref{sec:coupler_supp}.

Together, these results show that LIMINAL can treat hidden degrees of freedom as testable model extensions rather than fixed assumptions. By embedding latent subsystems into the Hilbert space and comparing nested models with likelihood-ratio tests, the framework distinguishes reduced effective descriptions from models that require explicit hidden-qubit dynamics or higher-locality interactions. This provides a common route for assessing couplers, unwanted two-level systems, and spectator qubits within the same model-selection workflow.

\section{Discussion}\label{sec:discussion}
We have introduced \emph{LIMINAL}, a data-driven framework for selecting \emph{minimal adequate} dynamical models of quantum processors by combining differentiable maximum-likelihood estimation with likelihood-ratio comparisons between nested Lindblad model families. The central purpose of the LIMINAL framework is to determine which physical processes are warranted by the data and where further increases in model complexity cease to provide explanatory power. Across the examples studied here, LIMINAL supports locality and connectivity selection, Hamiltonian reconstruction, and tests for hidden degrees of freedom within a common likelihood-based procedure.

We illustrate these capabilities in three complementary settings. For 5-qubit idling dynamics, the framework identifies a compact locality-structured model and detects saturation of explanatory power as the hierarchy is enlarged. For driven single-qubit experiments, it recovers physically interpretable Hamiltonian parameters for both constant and time-dependent control fields, including reconstruction of a shaped pulse directly at the Hamiltonian level. For coupler-mediated interactions, it provides a direct test of whether an effective reduced description is sufficient or whether hidden subsystems must be introduced explicitly. Taken together, these results show that the framework can both validate reduced descriptions and diagnose when additional structure is required by the data.

More generally, the selected adequate model should be interpreted relative to the hierarchy of extensions that is tested. In the present work, we varied locality and connectivity while fixing other assumptions, including restriction to the computational subspace and a specific SPAM description. It is therefore possible that alternative extensions, such as leakage-sensitive models, richer SPAM parameterizations, or explicitly non-Markovian dynamics, could account for part of the observed likelihood improvement more efficiently than further increases in locality. In that sense, the present analysis may be viewed as placing an upper bound on the locality required within the Lindblad model family considered here, rather than identifying a unique underlying physical description.

More broadly, the LIMINAL framework is not intended to remove the exponential cost of unconstrained Lindblad identification, but to justify compact structure before applying more specialized or compressed estimators~\cite{stilckfrancaEfficientRobustEstimation2024, birkeDemonstratingBenchmarkingClassical2026}. In that role, it can also be implemented adaptively: After fitting a baseline model, subsequent measurements could be designed to maximize sensitivity to the parameters that distinguish competing extensions, thereby reducing both experimental overhead and fitting cost.

\section{Methods}
\subsection{Gradient-based likelihood estimation of an ODE model}
\label{sec:gradient}

We estimate the parameters of a Lindblad model by maximizing the log-likelihood of time-resolved tomographic data under a differentiable, end-to-end simulation of the experiment. Let
$\Theta = \{\theta_\rho,\theta_H,\theta_L\}$ denote the parameters of the local initial state, the locality- and graph-structured Hamiltonian, and the locality- and graph-structured dissipator, respectively (cf.~Sec.~\ref{sec:param}). Given $\Theta$, the simulator maps each experimental configuration and duration to predicted measurement probabilities, and we optimize $\Theta$ via gradient-based updates using automatic differentiation combined with adjoint techniques for memory efficiency and second-order information when required (see practical implementation details in supplemental material~\ref{sec:practical}) \cite{kidgerNeuralDifferentialEquations2022b}.

Following the experimental protocol introduced in Sec.~\ref{sec:concept}, each experiment consists of (i) an \emph{initialization gate} \(U_i\) that prepares an informationally complete set of local states, (ii) evolution for a duration \(t\in T=\{0,\Delta t,\ldots,n_t\Delta t\}\) which is modeled by the Lindblad generator \(\mathcal{L}(\theta_H,\theta_L)\), and (iii) a \emph{measurement basis-change} \(U_m\) followed by projective  in the computational basis. We index \emph{configurations} by \(c\equiv(i,m)\), \emph{durations} by \(t\in T\), and \emph{measured bit strings} by \(b\in\{0,1\}^N\), corresponding to the \(2^N\) computational-basis bit strings of the \(N\)-qubit system. For each pair \((c,t)\), the observed counts are collected in the vector
\[
\mathbf{x}_{c,t}=\bigl(x_{c,t,b}\bigr)_{b\in\{0,1\}^N},
\]
where \(x_{c,t,b}\) is the number of times the bit string \(b\) is observed.

For parameters \(\Theta\), the simulator returns the corresponding probability vector
\[
\mathbf{p}_{c,t}(\Theta)=\bigl(p_{c,t,b}(\Theta)\bigr)_{b\in\{0,1\}^N}.
\]
The initial state for configuration \(c=(i,m)\) is
\begin{equation}
\rho_c(0;\theta_\rho)=U_i\,\rho_0(\theta_\rho)\,U_i^\dagger,
\end{equation}
where \(\rho_0(\theta_\rho)\) is the tensor-product initial state (Sec.~\ref{sec:param}). The time-evolved state \(\rho_c(t;\theta_H,\theta_L)\) is obtained by solving
\[
\dot{\rho}=\mathcal{L}(\theta_H,\theta_L)[\rho].
\]
The predicted probability of observing bit string \(b\) is then given by Born's rule after application of the measurement basis-change gate,
\begin{equation}
p_{c,t,b}(\Theta)
=
\bigl\langle b \big| U_m\, \rho_c(t;\theta_H,\theta_L)\, U_m^\dagger \big| b \bigr\rangle.
\end{equation}

For fixed \((c,t)\), each shot yields exactly one of the \(2^N\) possible bit-string outcomes, and the count vector \(\mathbf{x}_{c,t}\) is therefore modeled as multinomially distributed with outcome probabilities \(\mathbf{p}_{c,t}(\Theta)\) and total number of shots for each configuration and timestep \(N_{\mathrm{shots}}=\sum_b x_{c,t,b}\). Under the assumption of independent shots across repeated measurements, the full-data log-likelihood is
\begin{equation}
\mathrm{LL}(\Theta)
=
\sum_c \sum_{t\in T}
\log \mathrm{Multinomial}\!\left(\mathbf{x}_{c,t}\,;\,\mathbf{p}_{c,t}(\Theta),\,N_{\mathrm{shots}}\right).
\end{equation}
We estimate the parameters by maximum likelihood,
\[
\hat{\Theta}=\arg\max_\Theta \mathrm{LL}(\Theta).
\]

Algorithm~\ref{alg:gradient} summarizes the estimation procedure. Starting from either a random initialization or a warm start from a previously fitted nested model, we simulate the experiment for the current parameters, evaluate the multinomial log-likelihood, differentiate it with respect to \(\Theta\) through the ODE solver, and update the parameters using a gradient-based optimizer. In this way, the full forward model---state preparation, Lindblad evolution, basis change, and readout probabilities---remains differentiable end-to-end. Gradients are computed by automatic differentiation augmented by adjoint methods for memory efficiency and, when required, Hessian extraction (see Supplemental Material~\ref{sec:practical} for practical implementation details).

\begin{algorithm}[H]
\caption{Gradient-based maximum-likelihood estimation for Lindblad models}
\label{alg:gradient}
\begin{algorithmic}[1]
\State \textbf{Inputs:} data \(\mathbf{x}_{c,t}\); gates \(U_i,U_m\); durations \(T\); parameters \(\Theta=\{\theta_\rho,\theta_H,\theta_L\}\)
\State \textbf{Initialize:} sample each component of \(\Theta^{(0)}\) from \(\mathcal{N}(0,\sigma^2)\) or initialize from a fitted nested model
\Repeat
  \State \textbf{Forward pass:} compute \(\mathbf{p}_{c,t}(\Theta^{(i)}) \gets \textsc{Model}(\Theta^{(i)})\)
  \State \textbf{Objective:} \(\mathrm{LL}(\Theta^{(i)}) \gets \sum_{c,t} \log \mathrm{Multinomial}\!\bigl(\mathbf{x}_{c,t};\,\mathbf{p}_{c,t}(\Theta^{(i)}),\,N_{c,t}\bigr)\)
  \State \textbf{Gradient:} compute \(\nabla_\Theta \mathrm{LL}(\Theta^{(i)})\) via AD + adjoint method
  \State \textbf{Update:} \(\Theta^{(i+1)} \gets \textsc{Opt}\!\left(\Theta^{(i)},\nabla_\Theta \mathrm{LL}(\Theta^{(i)})\right)\) \Comment{e.g., Adam}
\Until{stopping criterion is met}
\State \textbf{Return} \(\hat{\Theta}=\Theta^{(i_\star)}\) and \(\mathrm{LL}(\hat{\Theta})\) evaluated on the full dataset
\end{algorithmic}
\end{algorithm}

Initial parameter values are obtained either by independent Gaussian sampling or by warm-starting from a previously fitted nested model. For random initialization, the sampling variance is chosen to decrease with locality so that higher-order terms are initially suppressed. When fitting a nested sequence of models, overlapping parameters from the smaller model provide a natural initial guess for the larger one.

Since gradient-based optimization does not guarantee convergence to the global optimum, we apply practical diagnostics to reduce the chance of spurious local minima. First, for truly nested models, the maximized log-likelihood of the larger model cannot be lower than that of the smaller one; any violation of this monotonicity is treated as evidence of incomplete optimization and triggers additional optimization effort, such as more iterations or restarts. Second, we inspect representative configurations and time traces to rule out pathological solutions, for example rapid collapse to an almost maximally mixed state that contradicts the observed dynamics. Where possible, each larger model is initialized from the fitted parameters of the preceding model to improve convergence. Specific choices of ODE solvers, adjoint modes, numerical precision, minibatching strategy, optimizer hyperparameters, and stopping criteria are provided in Supplemental Material~\ref{sec:practical}.

\subsection{Parameterizations for Lindblad models}
\label{sec:param}

Here we describe the parameterizations used to build the differentiable Lindblad models employed throughout this work. Following Sec.~\ref{sec:concept}, a \emph{parameterization} is a structured set of real variables, denoted by $\theta$, together with a differentiable map from those variables to a physical object, such as an initial state, a Hamiltonian, or a set of jump operators. A candidate model fixes the structural choices, such as locality and connectivity, and combines the corresponding parameterizations into \(\Theta=\{\theta_\rho,\theta_H,\theta_L\}\), where $\theta_\rho$ parameterizes the initial state, $\theta_H$ the Hamiltonian, and $\theta_L$ the dissipator. Together with the experimental sequence, this defines the differentiable map from $\Theta$ to simulated density matrices, predicted measurement probabilities, and hence the likelihood value used for fitting and model comparison (cf.~Sec.~\ref{sec:gradient}).
 
First, we parameterize the initial state. We use a local parameterization that builds the full $N$-qubit state as a tensor product of single-qubit density matrices. For each qubit $q\in\{1,\dots,N\}$ we introduce a real parameter matrix
\[
\theta_{\rho}^{\{q\}} \in \mathbb{R}^{2\times 2},
\]
which is mapped to a valid single-qubit density matrix $\rho^{\{q\}}$.
To guarantee positivity and differentiability, we parameterize the density matrix through its Cholesky-like factor. We directly construct a complex lower-triangular matrix
\[
B^{\{q\}}
=
\mathrm{tril}\!\bigl(\theta_{\rho}^{\{q\}}\bigr)
+
i\,\mathrm{triu}^{\!\circ}\!\bigl(\theta_{\rho}^{\{q\}}\bigr)^{T},
\]
and define
\[
\rho^{\{q\}}
=
\frac{B^{\{q\}} B^{\{q\}\dagger}}
{\Tr\!\left(B^{\{q\}} B^{\{q\}\dagger}\right)}.
\]
This construction ensures a positive semi-definite, trace-one state. The full initial state is then
\[
\rho(0;\theta_\rho)=\bigotimes_{q=1}^{N}\rho^{\{q\}}.
\]
While this introduces four real parameters per qubit—more than the three physical degrees of freedom—it maintains a smooth, unconstrained optimization landscape.

We next parameterize the Lindblad generator by specifying structured families for the Hamiltonian and dissipator. The structure is determined by a choice of locality and graph. For each $k\in\{1,\dots,k_{\max}\}$, let
\[
\mathcal{C}_k \subseteq \bigl\{\, (q_1,\dots,q_k)\; \big|\; 1\le q_1 < \cdots < q_k \le N \,\bigr\}
\]
denote the set of allowed $k$-qubit connections (hyperedges) in the interaction graph.
Each index $c\in\mathcal{C}_k$ therefore labels one specific $k$-local subsystem on which Hamiltonian or dissipative parameters may act. Different model families correspond to different choices of the connection sets \(\mathcal{C}_k\), for example local, nearest-neighbor, or all-to-all connectivity.

We expand the Hamiltonian in a locality hierarchy,
\[
H=\sum_{k=1}^{k_{\max}} H^{(k)},
\]
where \(H^{(k)}\) contains the terms supported on the connections in \(\mathcal{C}_k\). For one-local terms, we write
\begin{equation}
H^{(1)}
=
\sum_{q=1}^{N}\sum_{i\in\{x,y,z\}}
a^{(1)}_{q,i}\,\sigma_i^{\{q\}},
\label{eq:H1}
\end{equation}
where \(\sigma_i^{\{q\}}\) denotes the Pauli operator \(\sigma_i\) acting on qubit \(q\) and the identity elsewhere,
\[
\sigma_i^{\{q\}}
=
\mathbb{I}\otimes\cdots\otimes
\underbrace{\sigma_i}_{q\text{th position}}
\otimes\cdots\otimes\mathbb{I}.
\]
The coefficients are collected in the array \(a^{(1)}\in\mathbb{R}^{N\times 3}\). For two-local terms, we write
\begin{equation}
H^{(2)}
=
\sum_{c=(q,r)\in\mathcal{C}_2}
\sum_{i,j\in\{x,y,z\}}
a^{(2)}_{c,i,j}\,
\sigma_i^{\{q\}}\sigma_j^{\{r\}},
\label{eq:H2}
\end{equation}
with coefficients \(a^{(2)}\in\mathbb{R}^{|\mathcal{C}_2|\times 3\times 3}\). Higher-locality terms are defined analogously: for \(c=(q_1,\dots,q_k)\in\mathcal{C}_k\),
\[
H^{(k)}
=
\sum_{c\in\mathcal{C}_k}
\sum_{i_1,\dots,i_k\in\{x,y,z\}}
a^{(k)}_{c,i_1,\dots,i_k}
\prod_{\ell=1}^{k}\sigma_{i_\ell}^{\{q_\ell\}}.
\]
The Hamiltonian parameter set is therefore
\[
\theta_H=\bigl\{a^{(1)},a^{(2)},\dots,a^{(k_{\max})}\bigr\}.
\]
Every model of maximum locality \(k'<k_{\max}\) is obtained by setting all coefficients with locality greater than \(k'\) to zero. Provided \(k_{\max}\) remains fixed as \(N\) grows, this locality-based structure reduces the number of Hamiltonian parameters from exponential to polynomial in \(N\).

We now turn from the Hamiltonian terms to the dissipative part of the Lindblad generator. We parameterize dissipation through jump operators expanded in the Pauli basis, following Ref.~\cite{samachLindbladTomographySuperconducting2022a}. For a \(k\)-qubit subsystem, let
\[
\mathcal{P}_k
=
\bigl\{
P^{(k)}_b \;\big|\; b=1,\dots,4^k-1
\bigr\}
\]
denote the non-identity Pauli-string basis on \(k\) qubits. For a chosen maximum locality \(k\), each allowed connection \(c=(q_1,\dots,q_k)\in\mathcal{C}_k\) defines one dissipative block acting on that \(k\)-qubit subsystem. We introduce a real parameter matrix
\[
\theta_L^{(c)}\in\mathbb{R}^{(4^k-1)\times(4^k-1)},
\]
and map it to a complex lower-triangular coefficient matrix
\[
M^{(c)}
=
\mathrm{tril}\!\bigl(\theta_L^{(c)}\bigr)
+
i\,\mathrm{triu}^{\!\circ}\!\bigl(\theta_L^{(c)}\bigr)^T.
\]
Equivalently, \(M^{(c)}\) has real diagonal entries and unconstrained complex lower-triangular entries below the diagonal. This choice gives a differentiable real parameterization of a general complex lower-triangular matrix.

The block-local jump operators on the \(k\)-qubit subsystem \(c\) are then
\[
\widetilde{L}^{(c)}_a
=
\sum_{b=1}^{4^k-1}
M^{(c)}_{ab}\,P_b^{(k)},
\qquad
a=1,\dots,4^k-1.
\]
To embed these operators into the full \(N\)-qubit Hilbert space, we pad with identities and permute the active tensor factors onto the qubits specified by \(c\):
\[
L^{(c)}_a
=
\Pi_c
\left[
\widetilde{L}^{(c)}_a \otimes \mathbb{I}^{\otimes (N-k)}
\right]
\Pi_c^\dagger,
\]
where \(\Pi_c\) is the permutation operator that places the \(k\) active tensor factors on the qubit indices in \(c\). For a model with maximum dissipative locality \(k_{\max}\), the full set of jump operators is
\[
\bigl\{L_i^{(\le k_{\max})}\bigr\}
=
\bigcup_{c\in\mathcal{C}_{k_{\max}}}
\bigl\{L_a^{(c)}\bigr\}_{a=1}^{4^{k_{\max}}-1}.
\]
Since the basis \(\mathcal{P}_{k_{\max}}\) already contains Pauli strings with identities on subsets of the active qubits, each \(k_{\max}\)-local block also represents lower-locality dissipative terms on that subsystem.

\[
D^{(c)} = M^{(c)\dagger} M^{(c)},
\]
which contains the coefficients multiplying the Pauli strings acting from the left and right in the Lindblad dissipator (cf.\ Eq.~\ref{eq:lindblad_pauli}). By construction, $D^{(c)}$ is Hermitian and positive semidefinite. Redundancy can arise because lower-locality terms may be represented in more than one \(k\)-qubit block when several connections overlap. We allow this redundancy in the simulator and account for the number of \emph{independent real degrees of freedom} separately when performing likelihood-ratio tests. The full dissipative parameter set is denoted by
\[
\theta_L=\bigl\{\theta_L^{(c)} \;|\; c\in\mathcal{C}_{k_{\max}}\bigr\}.
\]
This yields a nested family of dissipative models ordered by both the chosen maximum locality and the underlying connection graph.

Taken together, the parameter sets \(\theta_H\) and \(\theta_L\) define nested model families ordered by maximum locality and by expansion of the allowed connection sets \(\mathcal{C}_k\) (for example from local to nearest-neighbor to all-to-all). Moving to a smaller model corresponds to fixing newly introduced parameters to zero, whereas moving to a larger model adds parameters that strictly contain those of the smaller one. This nested structure underpins the likelihood-ratio comparisons and explanatory-power analysis used throughout the paper. Specific details of the connection sets \(\mathcal{C}_k\), the implementation of the permutation operators \(\Pi_c\), and the bookkeeping of independent degrees of freedom for Wilks-type comparisons are given in Supplemental Material~\ref{sec:model_overview}.

\FloatBarrier
\begin{acknowledgments}
This research was supported by the Novo Nordisk Foundation (grant no. NNF22SA0081175), the NNF Quantum Computing Programme (NQCP), Villum Foundation through a Villum Young Investigator grant (grant no. 37467), the Innovation Fund Denmark (grant no. 2081-00013B, DanQ), the U.S. Army Research Office (grant no. W911NF-22-1-0042, NHyDTech), by the European Union through an ERC Starting Grant, (grant no. 101077479, NovADePro), and by the Carlsberg Foundation (grant no. CF21-0343). 
Any opinions, findings, conclusions or recommendations expressed in this material are those of the author(s) and do not necessarily reflect the views of Army Research Office or the US Government. 
Views and opinions expressed are those of the author(s) only and do not necessarily reflect those of the European Union or the European Research Council. Neither the European Union nor the granting authority can be held responsible for them. 
Finally, we gratefully acknowledge Lena Jacobsen and Helle Grunnet for program management support.
\end{acknowledgments}





\bibliography{references}


 






\title{Supplemental Material for: Model Testing of Superconducting Qubit Lindblad Dynamics by Gradient Optimization}

%


\appendix

\onecolumngrid
\section{Practical Implementation Details}
\label{sec:practical}

In this section, we provide additional details on the practical implementation of the classical analysis. All simulations and optimizations are implemented in Python using \emph{JAX}~\cite{bradburyJAXComposableTransformations2018} with just-in-time (JIT) compilation enabled for the end-to-end forward and backward passes. Ordinary differential equations (ODEs) are integrated with \emph{Diffrax}~\cite{kidgerNeuralDifferentialEquations2022b}. Multinomial log-likelihoods are evaluated with \emph{TensorFlow Probability}~\cite{martinabadiTensorFlowLargeScaleMachine2015}. Parameter updates use \emph{Optax}~\cite{deepmindDeepMindJAXEcosystem2020}. All experiments are run in 64-bit precision.

We integrate the Lindblad master equation with the Tsit5 solver~\cite{tsitourasRungeKuttaPairs2011} and built-in error control with absolute and relative tolerances of \(10^{-6}\). The solver outputs \(\rho_c(t;\Theta)\) at an application-defined set of \texttt{save-at} times \(T\) that coincide with the experimental truncation times.

Gradients of \(\mathrm{LL}(\Theta)\) are computed by automatic differentiation and adjoint calculation through the ODE solver. Depending on the experiment, we select among the following adjoint methods, all implemented in Diffrax:
\begin{enumerate}
    \item \textbf{Checkpointed Automatic Differentiation} (reverse-mode with checkpointing)~\cite{stummNewAlgorithmsOptimal2010, wangMinimalRepetitionDynamic2009}: used when memory permits, providing fast iteration speed. In this work, it is used for the hidden-qubit/coupler experiments of Sec.~\ref{sec:hidden_qubits}.
    \item \textbf{Backward adjoint}~\cite{chenNeuralOrdinaryDifferential2019a}: solves an auxiliary adjoint ODE for the gradient, reducing memory footprint. In this work, it is used for the five-qubit idling experiment of Sec.~\ref{sec:5qubit_idle}.
    \item \textbf{Direct adjoint}~\cite{kidgerNeuralDifferentialEquations2022b}: supports both forward and backward passes, enabling Hessian calculations needed for uncertainty estimation. In this work, it is used for the single-qubit constant-pulse and time-dependent pulse experiments of Secs.~\ref{sec:const} and~\ref{sec:gates}.
\end{enumerate}
The choice of adjoint mode affects runtime--memory trade-offs but not the model definitions in Sec.~\ref{sec:param}.

All parameters are initialized either from previously fitted models or independently from a zero-centered Gaussian distribution with fixed variance:
\[
\Theta^{(0)} \sim \mathcal{N}(0,\sigma^2)
\qquad
\text{for all components of }
\Theta=\{\theta_\rho,\theta_H,\theta_L\}.
\]
For random initialization, the variance is chosen to decrease with locality so that higher-order terms are initially suppressed. When fitting a nested sequence of models, overlapping parameters from the smaller model provide a natural initial guess for the larger one. 

To control memory use on large datasets, optimization steps are computed on minibatches of configurations and times. In the five-qubit idling experiment (Sec.~\ref{sec:5qubit_idle}), we use a minibatch fraction of \(1/16\) per parameter update. For model selection and all reported statistics, the final log-likelihood \(\mathrm{LL}(\hat{\Theta})\) is always evaluated on the full dataset. 

Parameter updates use the ADAM optimizer~\cite{kingmaAdamMethodStochastic2017a} with learning rates between \(10^{-4}\) for the five-qubit idling experiment and \(10^{-2}\) for the single-qubit reconstruction experiments of Secs.~\ref{sec:const} and~\ref{sec:gates}. Learning rates are kept constant throughout the fitting.

Optimization stops when at least one of the following conditions is met:
(i) the relative improvement in the batched-data log-likelihood falls below a small threshold for several successive evaluations (typically \(10^{-6}\) over one hundred iterations), or
(ii) a preset iteration limit is reached (typically around 5000 steps).

Likelihood-ratio tests compare nested models by the increase in log-likelihood versus the increase in \emph{independent real} parameters. We count degrees of freedom (DoF) as follows:
\begin{itemize}
    \item \textbf{Hamiltonian}: each coefficient \(a^{(k)}_{c,i_1,\dots,i_k}\in\mathbb{R}\) contributes one DoF. The total DoF equals the number of included Pauli strings on the allowed \(k\)-local connections \(\mathcal{C}_k\).
    \item \textbf{Dissipation}: for a \(k\)-local block with \(m=4^k-1\) non-identity Pauli strings, the lower-triangular complex matrix \(M^{(c)}\in\mathbb{C}^{m\times m}\) with real diagonal has \(m\) real diagonal DoF and \(m(m-1)\) real off-diagonal DoF, for a total of \(m^2\) real DoF (one-to-one with \(D^{(c)}=M^{(c)\dagger}M^{(c)}\)). When assembling all blocks across \(\mathcal{C}_k\), a local operator can appear in multiple \(k\)-local embeddings. We allow this redundancy in the simulator but count DoF only once per independent real parameter when computing the DoF difference \(k_{\mathrm{add}}\) for Wilks’ theorem.
\end{itemize}
\section{Measurement setup}\label{sec:setup}
We provide here a concise overview of the measurement infrastructure used throughout this work. The device used is a five qubit processor with fixed frequency transmons and 4 tunable couplers. An optical micrograph is shown in Fig.~\ref{fig:concept} with false colored qubits and couplers. The device was fabricated at Chalmers University of Technology.

\begin{figure}[H]
    \centering
    \includegraphics{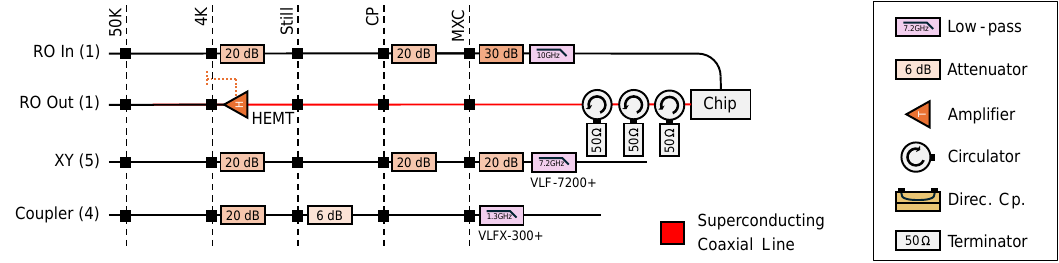}
    \caption{\textbf{Cryogenic measurement setup.} Schematic of the wiring used for the experiments, showing the XY (drive), Z (flux), readout input (RO~In) and readout output (RO~Out) lines across the cryostat stages (50\,K, 4\,K, Still/CP, MXC). The amplification chain comprises a TWPA at MXC followed by a HEMT amplifier at 4\,K; attenuators, filters and thermalization points are indicated at their respective stages. Control and acquisition are performed with an OPX1000 (HF~FEM for pulsed microwave control; LF~FEM for flux control).}
    \label{fig:setup}
\end{figure}

A schematic of the cryogenic wiring and amplification chain is shown in Fig.~\ref{fig:setup}. The device is mounted in a QCage.24 attached to the mixing-chamber (MXC) plate of a dilution refrigerator (Bluefors XLD1000). Control and readout are performed with a Quantum Machines OPX1000 controller equipped with high-frequency front-end modules (HF FEM) for pulsed microwave control and low-frequency front-end modules (LF FEM) for flux control of the tunable couplers. All qubit calibration, characterization and experiment orchestration are executed using our in-house software stack \textit{Pelagic}.%

Each qubit is addressed by a dedicated XY control line for microwave pulses and a Z (flux) line for static or dynamically modulated flux control of the couplers. The lines are routed through a series of cryogenic components (attenuators, filters and thermal anchors) distributed across the temperature stages (50\,K, 4\,K, Still/CP and MXC), as indicated in Fig.~\ref{fig:setup}.

Readout signals are delivered to the chip via a common input line and the emitted signals are routed to the output chain. The readout signal is amplified by a cryogenic high-electron-mobility-transistor (HEMT) amplifier at the 4\,K stage. Additional room-temperature amplification is performed and the signal is demodulated and digitized by the OPX1000. A full overview of the dilution refrigerator configuration is provided in Fig.~\ref{fig:setup}.%

\section{Qubit Characterization and Benchmarking}\label{sec:characterization-benchmarking}
We characterize the five qubits using standard microwave spectroscopy, time‑domain calibration, and randomized benchmarking protocols analogous to the offloading approach in \cite{marciniakMillisecondScaleCalibrationBenchmarking2026}. Qubit transition frequencies $f_{01}$ are located by spectroscopy; after identifying a $\pi$‑pulse via a Rabi experiment, the working frequencies are fine‑tuned with a Ramsey measurement. Energy‑relaxation and dephasing times ($T_{1}$, $T_{2}$) are obtained from conventional coherence experiments and are later compared with the parameters inferred by our modeling framework.

\begin{figure}[t]
    \centering
    \includegraphics[width=1.0\linewidth]{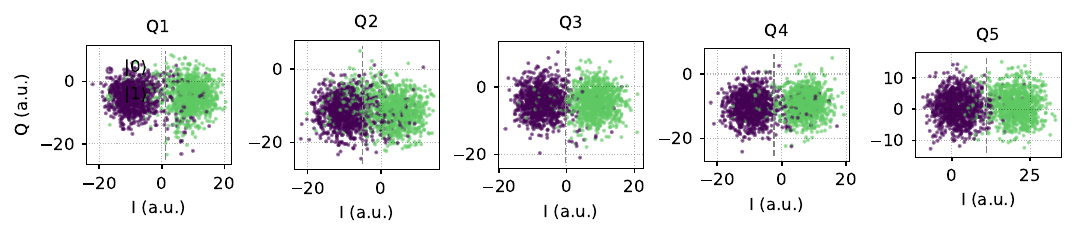}
    \caption{IQ data showing the distribution of single shot data for the five different qubits.}
    \label{fig:iq}
\end{figure}

Dispersive readout is calibrated against a high‑$Q$ resonator using a $\SI{6}{\micro s}$ constant readout pulse followed by an additional $\SI{2}{\micro s}$ demodulation window. We optimize the rotation angle, threshold and matched‑filter weights to maximize the single‑shot assignment fidelity
\[
F_{\mathrm{RO}} \equiv 1 - p(\text{``0''}\,|\,\ket{1}) - p(\text{``1''}\,|\,\ket{0}) \, .
\]
Representative IQ clouds (measured immediately prior to the idling experiment) are shown in Fig.~\ref{fig:iq}; the corresponding $F_{\mathrm{RO}}$ values are reported in Table~\ref{tab:qubit_characterization}.

\begin{table}[H]
\centering
\caption{Characterization of Qubits}
\label{tab:qubit_characterization}
\begin{tabular}{lcccccc}
\hline
Qubit & $f_{01}$ [GHz] & T1 [µs] & T2 [µs]  & $F_{\mathrm{RO}}$  & $\epsilon_g^{\mathrm{avg}}$ [\%] \\
\hline
Q1 & 3.59 & $190 \pm 4$ & $20 \pm 3$ & 0.843  & 0.06  \\
Q2 & 3.70 & $93  \pm 2$ & $80 \pm 20$ & 0.714 & 0.09  \\
Q3 & 3.78 & $95  \pm 3$ & $56 \pm 7$ & 0.921  & 0.06  \\
Q4 & 3.79 & $55  \pm 1$ & $37 \pm 3$ & 0.882  & 0.17  \\
Q5 & 4.12 & $90  \pm 2$ & $36 \pm 6$ & 0.916  & 0.06  \\
\hline
\end{tabular}
\end{table}

\begin{figure}
    \centering
    \includegraphics[width=1.0\linewidth]{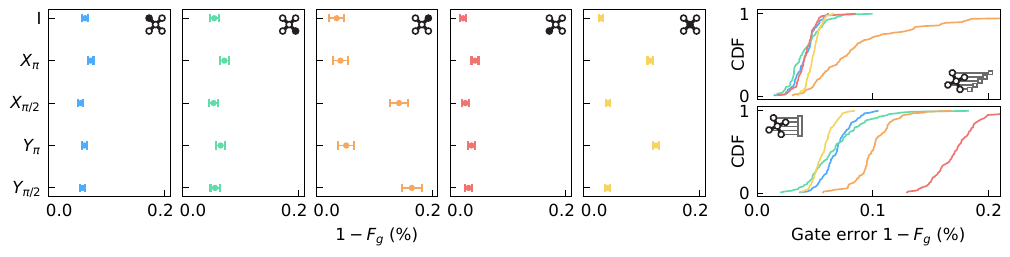}
    \caption{Fidelities for different operations in the gate set extracted via interleaved randomized benchmarking \cite{magesanEfficientMeasurementQuantum2012}. The right panel show standard Clifford randomized benchmarking \cite{magesanCharacterizingQuantumGates2012} run both sequentially (top)  and simulatanous (lower).}
    \label{fig:icrb}
    \includegraphics[width=1.0\linewidth]{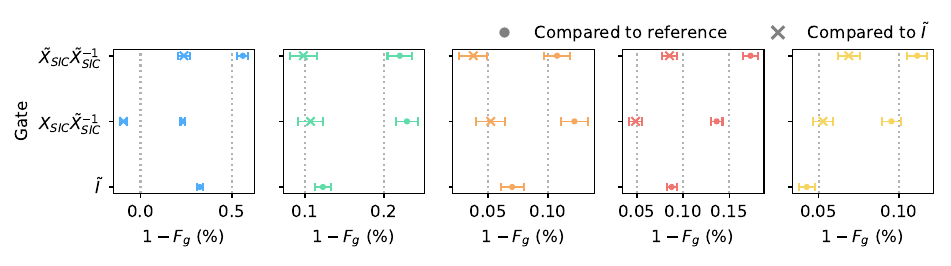}
    \caption{Fidelities for identities constructed from the $X_\mathrm{SIC}$ used to prepare in the tetrahedron states. We show the difference between decomposed SIC gate (top), SIC gate and decomposed inversion (middle), decomposed identity lower. We compare the gate fidelities both to the reference (no gates played) and to the identity to extract approximate fidelity of the $X_\mathrm{SIC}$ gate.}
    \label{fig:SIC_crb}
\end{figure}

Gates are calibrated iteratively after finding an initial pulse by applying a Rabi style experiment, we follow with error amplification \emph{train} to find a better amplitude. This is followed by DRAG experiment to add a Q quadrature optimizing the pulse \cite{motzoiSimplePulsesElimination2009a}. Gate calibration ends with individual pulse-train experiments to optimize for both 180 and 90 degree gates. 

We benchmark the native single‑qubit gates used for state preparation by Clifford randomized benchmarking (CRB), including both sequential and simultaneous CRB, and by interleaved CRB to estimate individual gate errors. Fig.~\ref{fig:icrb} summarizes interleaved‑RB fits for the preparation gates and compares sequential versus simultaneous operation; the average gate error extracted from standard CRB is listed in Table~\ref{tab:qubit_characterization}.

\paragraph{Tetrahedron (SIC) state preparation.}
To prepare the four tetrahedron states used in section \ref{sec:5qubit_idle}, we make use of an $2\arctan\!\sqrt{2}$ rotation around X and follow with virtual‑$Z$ rotations by $0$, $2\pi/3$, or $4\pi/3$. This allow us to prepare the following states:
\[
\ket{\psi} \in \Bigl\{\ket{0},\ \sqrt{\tfrac{1}{3}}\ket{0} + i\sqrt{\tfrac{2}{3}}\ket{1},\ \sqrt{\tfrac{1}{3}}\ket{0} + i e^{i2\pi/3}\sqrt{\tfrac{2}{3}}\ket{1},\ \sqrt{\tfrac{1}{3}}\ket{0} + i e^{i4\pi/3}\sqrt{\tfrac{2}{3}}\ket{1}\Bigr\},
\]
To perform benchmarking of the gate, we perform interleaved benchmarking of identities, which are decomposed into combinations of the $X_{\mathrm{SIC}}$ and its decomposition into $X_{\pi/2}$ and virtual Z rotations \cite{mckayEfficient$Z$Gates2017}. Virtual‑$Z$ gates are treated as ideal frame updates. We perform three interleaved experiments and a Clifford randomized benchmarking for a reference.

First, we will interleave a decomposed identity to get a reference for the decomposed version of an arbitrary SU(2) rotation. We interleave:
\begin{equation}
    \tilde{\mathbb{I}} = Z_{-\pi/2} X_{\pi/2} Z_{\pi} X_{\pi/2} Z_{-\pi/2},
\end{equation}
Secondly, we perform interleaved randomized benchmarking, where we decompose the $X_\mathrm{SIC}$ gate and decompose its inverse to get the identity. We both provide the error of this identity when compared with the reference CRB and when we compare to the interleaved decomposition done above. 
\begin{equation}
    \mathbb{I} = Z_{-\pi/2} X_{\pi/2} Z_{-\theta-\pi} X_{\pi/2} Z_{-\pi/2} \, Z_{-\pi/2} X_{\pi/2} Z_{+\theta-\pi} X_{\pi/2} Z_{-\pi/2},
\end{equation}

Finally, we apply the one-pulse $X_\mathrm{SIC}$ followed by its decomposed inverse. 
\begin{equation}
    \mathbb{I} = Z_{-\pi/2} X_{\pi/2} Z_{-\theta-\pi} X_{\pi/2} Z_{-\pi/2} \, X_{\mathrm{SIC}},
\end{equation}

We see that the errors associated with the one-pulse gate matches or exceeds the fidelity of the decomposed version. We find that playing the $X_\mathrm{SIC}$ gate with one pulse has the better performance and we extract infidelity of the gate of order $1-F_G\approx0.1 \%$, see Fig.~\ref{fig:SIC_crb}.

\FloatBarrier
\section{Overview over Experiments}\label{sec:ëxp_overview}
This section provides an overview over the experiments performed for this work. In Table \ref{tab:data_summary}, we reference the number of shots, initialization states, measurement bases and wall clock time for experiments. To reduce the time for each experiment, we employ an active reset protocol where every qubit is measured and a conditional $X$-gate is applied if the measured qubit was in its excited state. This is repeated until the qubit is measured in the ground state. The qubits are initialized when all involved qubits part are measured in their ground state. 

\begin{table}[H]
\centering
\caption{Data set sizes and acquisition parameters.}
\label{tab:data_summary}
\begin{tabular}{lcccccc}
\hline
Experiment & $N_{\text{shots}}$ & preparations & bases  & $t's$ range (step)  & Total shots & Duration \\
\hline
5-qubit idle & 100 & $4^5$ (SIC)                        & $3^5$ (x, y, z)   & $\SI{30}{\micro s}$ ($\SI{1}{\micro s}$) & $7.71\times10^8$  & 14 h   \\
1-qubit const pulse & 100 & $6$ (Pauli)                 & $3$ (x, y, z)  & $\SI{4.9}{\micro s}$ ($\SI{100}{ns}$) + 41 amplitudes &  $3.69\times10^6$ &  3 min\\
1-qubit $H(t)$ & 1000 & $6$ (Pauli)                       & $3$ (x, y, z)     & $\SI{120}{ns}$ ($\SI{4}{ns}$) &  $5.76
\times10^5$ & 25 s  \\
Coupler experiment & 25    & $6^2$ (Pauli) & $3^2$ (x, y, z)     & $\SI{3.92}{\micro s}$ ($\SI{80}{ns}$) &  $3.96\times10^5$ & 20 s \\
\hline
\end{tabular}
\end{table}

\section{Overview over Models}\label{sec:model_overview}
In this section, we provide an overview of the models used for fitting the tomographic data through the main work. For any time-independent dynamics, Hamiltonians have \#dof equal to $3^k$ where $k$ is the locality for each connection of that locality. For a Hamiltonian of $k_\text{max}$-locality we sum all the contributions from smaller small to large. For a dissipation parameterization describing a $k_{\text{max}}$-local interaction, the degrees of freedom are  $(4^{k_{\text{max}}}-1)^2$. With more interactions we sum the number from each connection but have to remove double counted interaction of smaller subsets 

\paragraph{Idling experiment} Here the models include a parameterization for a Lindblad generator for simultaneous five qubit dynamics. Models are build by combining one of five Hamiltonian and dissipative graph structures. The number of parameters for each are collected in the table below.

\begin{table}[H]
\centering
\caption{Locality‑structured model families for the five‑qubit idling experiment. ``Connections'' lists the number of $k$‑qubit subsets in the specified graph (five single‑qubit sites; four nearest‑neighbor pairs; all (ten) pairs; ten three qubit subsets). Degrees of freedom (DoF) for Hamiltonian and dissipator follow the parameterizations used in the main text and SI.}
\label{tab:idling_models}
\begin{tabular}{lcccccc}
\hline
\textbf{Name} & \textbf{Max locality} & \textbf{Connections} & \textbf{Hamiltonian DoF} & \textbf{Dissipator DoF} \\
\hline
none        & 0 & 0  & 0   & 0      \\
local       & 1 & 5  & 15  & 45     \\
nearest     & 2 & 4  & 51  & 855    \\
all pairs   & 2 & 10 & 105 & 2025   \\
3‑local     & 3 & 10 & 375 & 32940  \\
\hline
\end{tabular}
\end{table}

\paragraph{Single Qubit Experiments}
For the constant pulse reconstruction each amplitude has its own model, all consisting of a time-independent Hamiltonian and dissipation matrix giving a total of 15 parameters in addition to single qubit state preparation errors. The time-dependent envelop is modeled with a 15 Hamiltonians equally distributed over the pulse duration. The gives a total of 45 parameters for Hamiltonian and 9 for the dissipation in addition state preparation errors. 

\paragraph{Hidden Coupler Dynamics}
For the hidden coupler dynamics, the models used follows the same patterns as for the idling dynamics. However, for any hidden qubit, we will have a unitary degree of freedom of our qubit. Meaning that we in practice have three less parameters per hidden qubit. An overview of the models are provided in Table \ref{tab:coupler_models}.

\begin{table*}[h]
\centering
\caption{Locality‑structured model families for coupler‑mediated dynamics (Sec.~II\,E). 
``Observed tomography'' indicates whether measurements are performed on one or two qubits; hidden subsystems are traced out prior to likelihood evaluation. 
``Pairs'' lists the number of two‑qubit connections in the chosen interaction graph for the dissipator (two‑local dissipation is included whenever two‑qubit interactions are present). 
Hamiltonian and dissipator DoF follow the parameterizations in the Methods/SI. Total DOF are the combined count for Hamiltonian and dissipation with 3 subtracted for the unitary freedom per hidden qubit.}
\label{tab:coupler_models}
\begin{tabular}{lccccccc}
\hline
\textbf{Name} & \textbf{Observed Qubits} & \textbf{Hidden qubits} & \textbf{Max locality (H)} & \textbf{Pairs} & \textbf{H DoF} & \textbf{D DoF} & \textbf{Total DoF}\\
\hline
A: single‑qubit effective model     & 1 & 0 & 1 (local)                  & 0 & $3$   & $9$   & 12  \\
B: A + one hidden TLS               & 1 & 1 & 2 (up to pairs)            & 1 & $15$ & $225$  & 237 \\
C: B + two hidden TLS               & 1 & 2 & 2 (up to pairs)            & 3 & $36$ & $630$  & 660 \\
\hline
D: two‑qubit effective model        & 2 & 0 & 2 (up to pairs)            & 1 & $15$ & $225$  & 240 \\
E: D + hidden couple                & 2 & 1 & 2 (up to pairs)            & 2 & $36$ & $630$  & 653 \\
F: E + 3‑local H                    & 2 & 1 & 3 (add 3-local term)        & 3 & $63$ & $630$  & 690 \\
\hline
\end{tabular}
\end{table*}

\section{Additional Data on Idling Experiment}\label{sec:idling_supp}
\begin{figure}
    \centering    \includegraphics{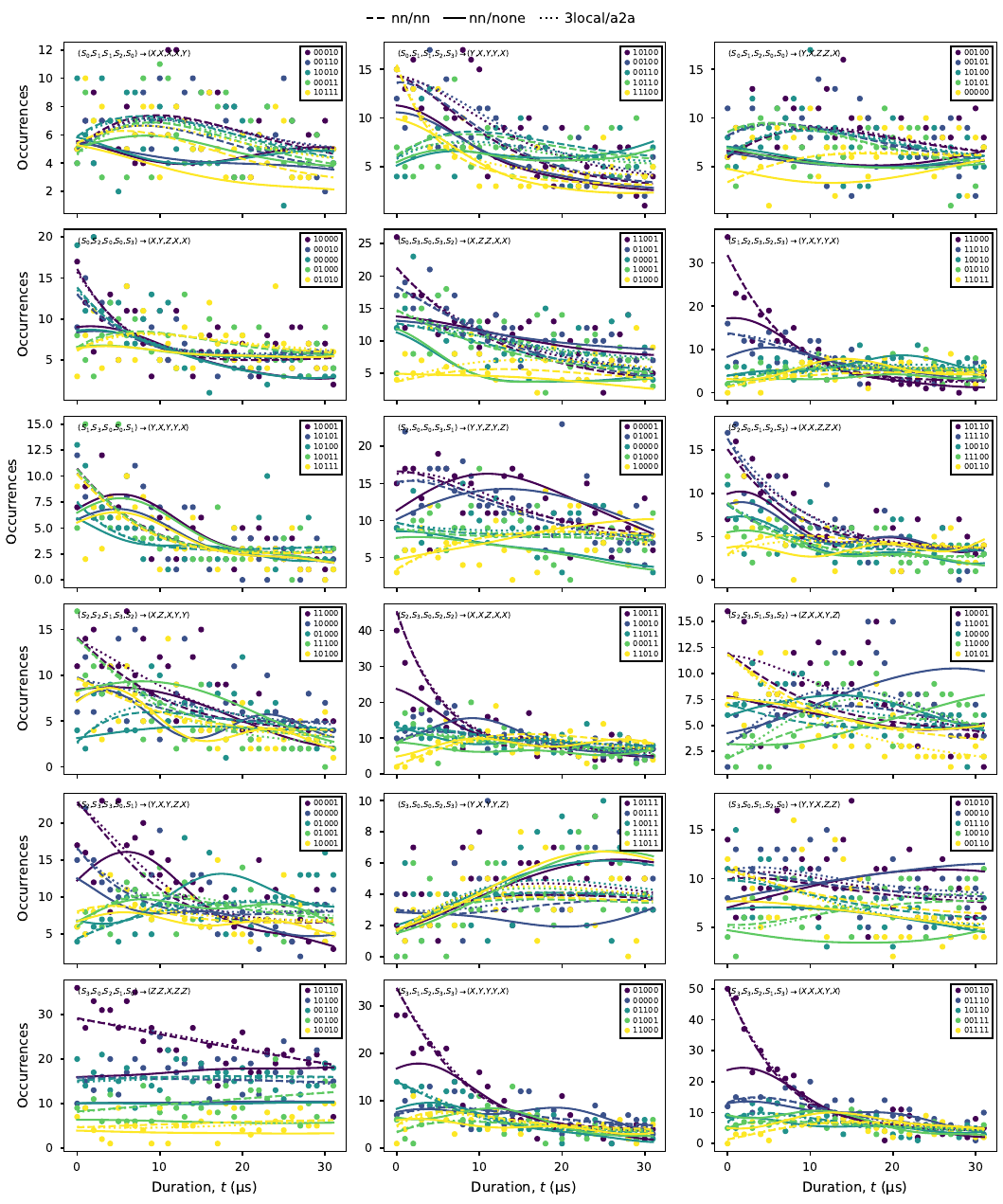}
    \caption{Random selection of 18 configurations out of 248,832 different tomographical experiments. Three models are visualized, the neareast neighbor interaction and no dissipation, the near neighbor interaction and nearest neight dissipation and the all-to-all interaction with all-to-all dissipation. For each configuration the 5 outcomes with the highest average population over the entire duration is displayed.}
    \label{fig:idling_supplementary}
\end{figure}

We summarize additional information about the idling experiment from main text section IIB here. The dataset consist of 248,832 different tomographic configuration for which we have fitted 25 different models. In addition to the configuration displayed in Fig.~\ref{fig:model_selection}C, we here provide 18 additional experimental configurations in Fig.~\ref{fig:idling_supplementary}. For each trace we show the simulation with the estimated parameters for three different models. We show the nearest neighbor Hamiltonian and no dissipation (full line), nearest neighbor dissipation and Hamiltonian (dashed) and finally the all pair dissipation and Hamiltonian dotted. For each configuration, we have selected the five bit strings which over the entire duration has the highest average number of occurrences.

In Table \ref{tab:five_qubit_models}, we provide the final negative log likelihood value each of the 25 fitted models, number of parameters as well as Akaike Information Criterion (AIC) and Bayesian Information Criterion (BIC). Which overall agrees with the findings of following the Explanatory power path used in section \ref{sec:5qubit_idle}. We see here that the choice of path leaves some interpretation whether one should include three-local Hamiltonian. If we were to start from the most complex model and do the remove the complexity which reduces the explanatory power by the least, we would end with the all pair dissipation and Hamiltonian. This is also the result if we were to compare the AIC or BIC criterion.

\begin{table}
\centering
\caption{Comparison of five-qubit Lindbladian models sorted by negative log-likelihood (NLL). $d$ denotes the number of free parameters. AIC $= 2\,\mathrm{NLL} + 2d$, BIC $= 2\,\mathrm{NLL} + d \ln N_\mathrm{obs}$ with $N_\mathrm{obs} = 796,262,400$.}
\label{tab:five_qubit_models}
\begin{tabular}{ll rr rr}
\hline
Hamiltonian & Jump Op. & NLL ($10^6$)  & $d$ & AIC($10^6$) & BIC ($10^6$) \\
\hline
3local & 3local & 407.90 & 35,430 & 815.86 & 816.52 \\
a2a & 3local & 407.90 & 35,160 & 815.86 & 816.51 \\
a2a & a2a & 407.90 & 2,220 & 815.80 & 815.84 \\
3local & a2a & 407.91 & 2,490 & 815.82 & 815.87 \\
a2a & nn & 407.97 & 978 & 815.94 & 815.96 \\
3local & nn & 407.99 & 1,248 & 815.99 & 816.01 \\
3local & local & 408.19 & 420 & 816.38 & 816.39 \\
nn & 3local & 408.20 & 35,106 & 816.47 & 817.12 \\
nn & a2a & 408.62 & 2,166 & 817.24 & 817.28 \\
local & 3local & 409.26 & 35,070 & 818.58 & 819.23 \\
a2a & local & 409.30 & 150 & 818.61 & 818.61 \\
local & a2a & 410.00 & 2,130 & 820.00 & 820.04 \\
none & 3local & 410.76 & 35,055 & 821.59 & 822.24 \\
none & a2a & 412.20 & 2,115 & 824.40 & 824.43 \\
nn & nn & 414.30 & 924 & 828.60 & 828.62 \\
nn & local & 414.35 & 96 & 828.70 & 828.71 \\
local & nn & 416.46 & 888 & 832.93 & 832.95 \\
local & local & 416.68 & 60 & 833.37 & 833.37 \\
none & nn & 418.46 & 873 & 836.92 & 836.93 \\
none & local & 426.67 & 45 & 853.35 & 853.35 \\
3local & none & 449.55 & 375 & 899.10 & 899.11 \\
a2a & none & 450.25 & 105 & 900.50 & 900.50 \\
nn & none & 470.39 & 51 & 940.79 & 940.79 \\
local & none & 488.90 & 15 & 977.79 & 977.79 \\
none & none & 512.49 & 0 & 1024.99 & 1024.99 \\
\hline
\end{tabular}
\end{table}

\FloatBarrier
\section{Additional data on the constant pulse experiment}\label{sec:const_supp}

This section complements the constant, slightly detuned pulse benchmark in the main text (Sec.~II\,C) by showing the full set of configurations and the fitted dissipators. For each drive amplitude $A$, we acquire tomographic data for six initial states and three measurement axes, yielding $18$ configurations that are \emph{jointly} fitted by maximum-likelihood estimation (MLE) under a single-qubit, time-independent Lindblad generator. Successive amplitudes are warm-started from the previous fit to promote continuity across $A$. See Sec.~II\,C for the experiment description and fitting protocol (MLE through the ODE-based simulator), including the use of warm starts across amplitudes.%

\begin{figure}
    \centering  \includegraphics{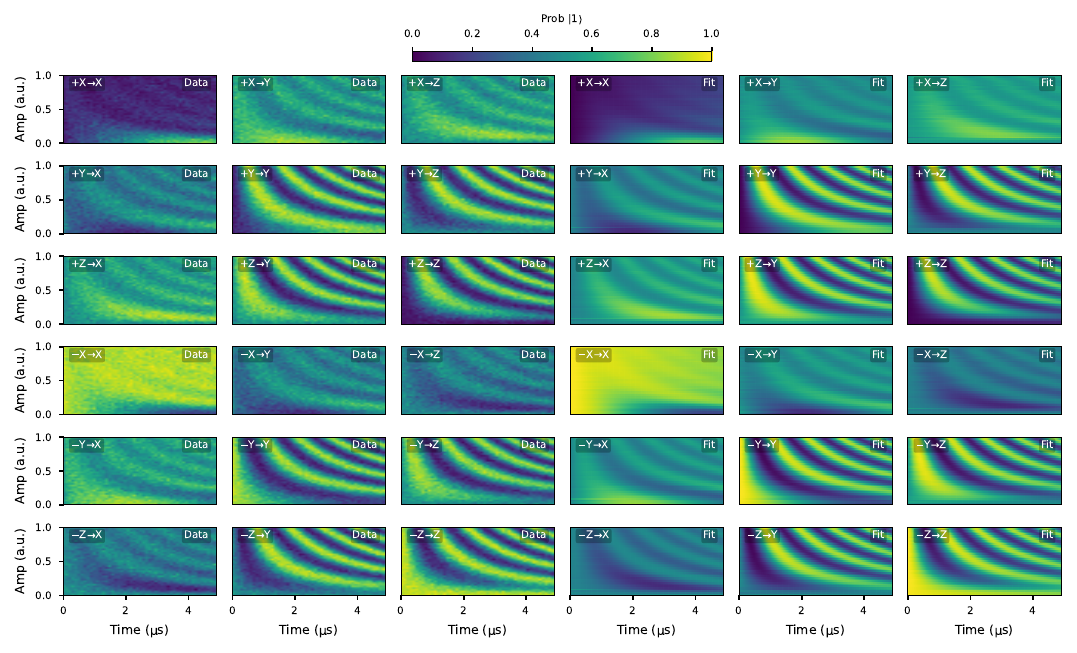}
    \caption{\textbf{Constant-pulse benchmark: full configurations and fits.} 
    Left: measured probabilities for six initial states and three measurement axes (18 configurations) at multiple amplitudes $A$. 
    Right: simulations from the MLE parameters of a single-qubit, time-independent Lindblad model at each $A$.}
    \label{fig:fits_for_const}
\end{figure}

\paragraph{Full dataset and fits.}
Fig.~\ref{fig:fits_for_const} displays the complete dataset for all amplitudes and all $18$ configurations (left three columns) together with the corresponding simulations generated from the MLE parameter estimates (right three columns). The fixed single-qubit, time-independent model at each $A$ consists of a Hamiltonian expanded in the Pauli basis and a $3\times3$ dissipator acting on the non-identity Pauli components, as detailed in the Methods.

\begin{figure}[H]
    \centering
    \includegraphics{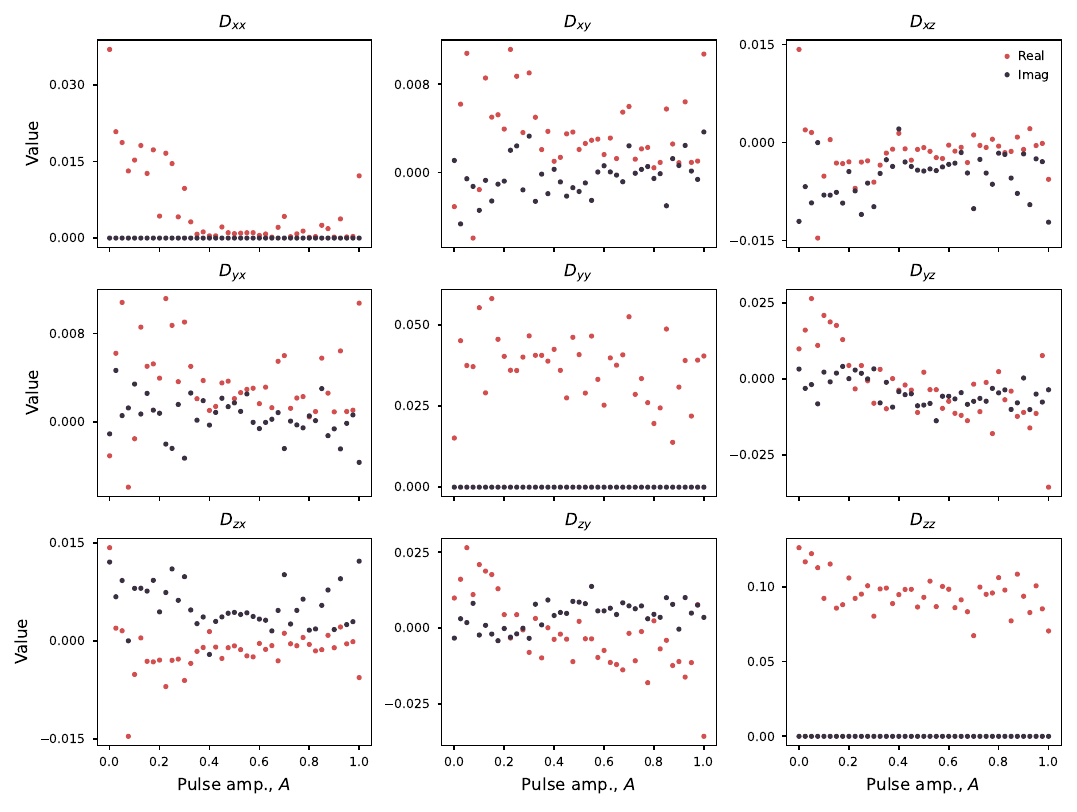}
    \caption{\textbf{Amplitude dependence of the dissipator.} 
    Components of the dissipator $D_{mn}$ extracted at each amplitude $A$ for the constant-pulse experiment (Sec.~II\,C). 
    By construction $D$ is Hermitian and positive semi-definite; each element $D_{mn}$ weighs the Lindblad term with $\sigma_m$ acting from the left and $\sigma_n$ from the right, cf.~Eq.~\eqref{eq:lindblad_pauli_const}.}
    \label{fig:dissipation_for_const_pulse}
\end{figure}

\paragraph{Estimated dissipators.}
For each amplitude $A$, the dissipative part of the generator is represented by a Hermitian, positive semidefinite $3\times3$ matrix $D$ over the Pauli basis $\{\sigma_x,\sigma_y,\sigma_z\}$ (excluding the identity), so that the full one-qubit Lindblad master equation reads
\begin{align}
    \frac{d\rho}{dt} 
    &= -\frac{i}{\hbar}\sum_{i\in\{x,y,z\}} a_i\,[\sigma_i,\rho]
     \;+\; \sum_{m,n\in\{x,y,z\}} D_{mn}\!\left(\sigma_m \rho\, \sigma_n - \tfrac12\{\sigma_n\sigma_m,\rho\}\right).
    \label{eq:lindblad_pauli_const}
\end{align}
This Pauli-basis formulation matches the parameterization used throughout the paper, where $D=M^\dagger M$ by construction, ensuring $D\succeq 0$.%
Fig.~\ref{fig:dissipation_for_const_pulse} shows the recovered $D_{mn}(A)$ components for all fitted amplitudes.

\section{Additional Data on Enveloped Pulse Experiment}\label{sec:gate_supp}
\begin{figure}
    \centering
\includegraphics{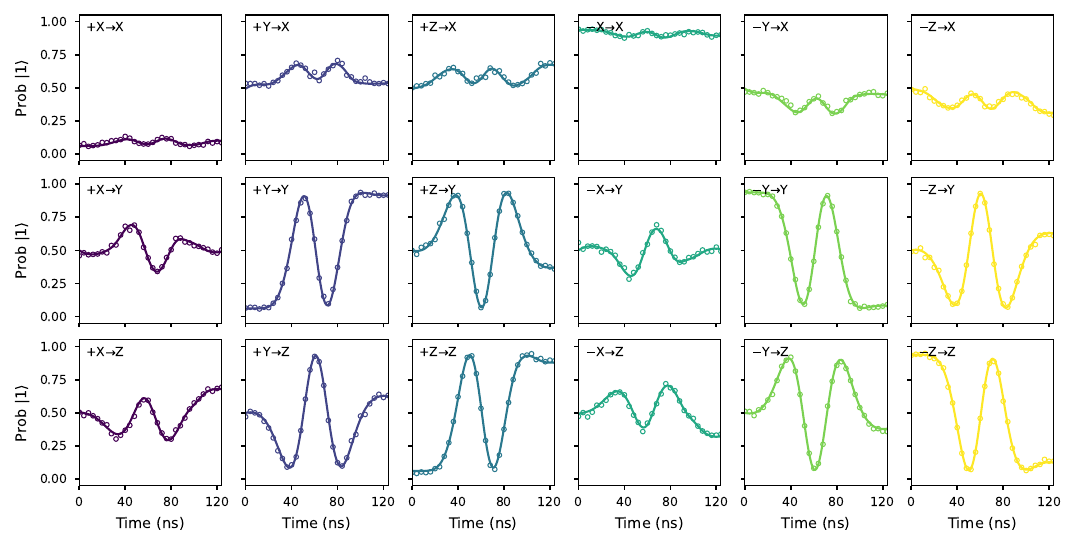}
    \caption{Overview of fit and data for all visualization of the enveloped function. The title refers to initialized state and the measurement basis.}
    \label{fig:time-dep-tomo}
\end{figure}

This section serves to compliment the experiment performed in the main text in section \ref{sec:gates}, where a time-dependent pulse was reconstructed. All 18 configurations are shown in figure \ref{fig:time-dep-tomo} along with the simulation based on the estimated parameters of the time-dependent Hamiltonian.

The data was fitted with time-independent dissipation. The dissipation matrix of the reconstruction was found to be:

\begin{align*}
    \mathbf{D} &=
    \begin{pmatrix}
     0.035 & 0.015 + 0.023\,\mathrm{i} & -0.0040 + 0.0056\,\mathrm{i} \\
     0.015 - 0.023\,\mathrm{i} & 0.032 & 0.0047 + 0.0051\,\mathrm{i} \\
     -0.0040 - 0.0056\,\mathrm{i} & 0.0047 - 0.0051\,\mathrm{i} & 0.0029
    \end{pmatrix} \unit{(\micro s)}^{-1}
\end{align*}
showing weak dissipation. Here the errors of the parameterization are large, and are not propagated to the D matrix, since the error landscape would be highly non-gaussian. With bootstrapping the D-matrix and its eigenvalues can be found, but the distributions are non-gaussian and the confidence region goes from around 0 to around 10 times the numerical value reported.

\section{Additional Data on the Coupler Experiment}\label{sec:coupler_supp}
\begin{figure}
    \centering
    \includegraphics{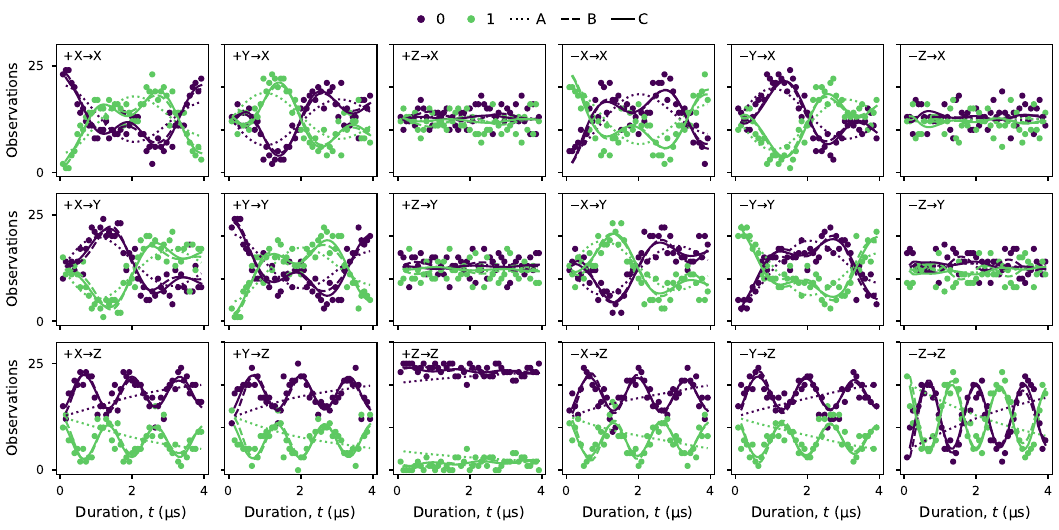}
    \caption{Overview of fit and data for all visualization of the 1 qubit tomography of the driven coupler experiment. The title refers to initialized state and the measurement basis.}
    \label{fig:1qubit-coupler-tomo}
\end{figure}

\begin{figure}
    \includegraphics{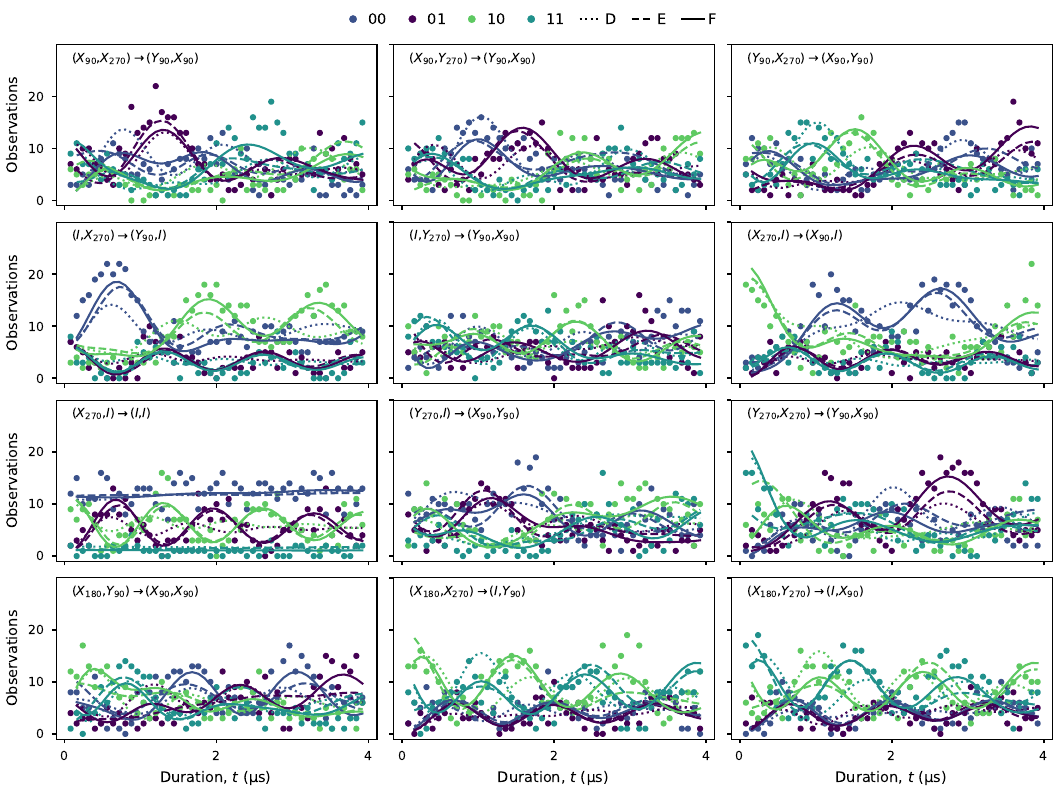}
    \caption{Overview of fit and data for 12 randomly selected configurations of the 2 qubit tomography of the driven coupler experiment. The title refers to the gates played before and after the coupler drive.}
    \label{fig:2qubit-coupler-tomo}
\end{figure}

This section compliments the experiment on hidden qubits by the coupler mediated two-qubit interaction. In addition to the data visualized in Fig.~\ref{fig:coupler}, we here provide the data and fit for all possible configurations for the one-qubit tomographic set (Fig.~\ref{fig:1qubit-coupler-tomo}). Of the 324 different configurations for the two qubit tomographic dataset, we provide 12 chosen at random in Fig.~\ref{fig:2qubit-coupler-tomo}.


\FloatBarrier




\end{document}